# Universal correlation between electronic factors and solute-defect interactions in bcc refractory metals

Yong-Jie Hu[1], Ge Zhao[2], Baiyu Zhang[3], Chaoming Yang[1], Zi-Kui Liu[4], Xiaofeng Qian[3], and Liang Qi[1]

[1]Department of Materials Science and Engineering, University of Michigan, Ann Arbor, Michigan 48109, USA

[2]Department of Statistics, Pennsylvania State University, State College, Pennsylvania 16802, USA

[3]Department of Materials Science and Engineering, Texas A&M University, College Station, Texas 77843, USA

[4]Department of Materials Science and Engineering, Pennsylvania State University, State College, Pennsylvania 16802, USA

**Abstract:**

The interactions between solute atoms and crystalline defects such as vacancies, dislocations, and grain boundaries play an essential role in determining physical, chemical and mechanical properties of solid-solution alloys. Here we present a universal correlation between two electronic factors and the solute-defect interaction energies in binary alloys of body-centered-cubic (bcc) refractory metals (such as W and Ta) with transition-metal substitutional solutes. One electronic factor is the bimodality of the *d*-orbital local density of states for a matrix atom at the substitutional site, and the other is related to the hybridization strength between the valance *sp*- and *d*-bands for the same matrix atom. Remarkably, the correlation is independent of the types of defects and the locations of substitutional sites, following a linear relation for a particular pair of solute-matrix elements. Our findings provide a novel and quantitative guidance to engineer the solute-defect interactions in alloys based on electronic structures.

Solute atoms, whether they are added voluntarily for specific needs, inevitably remained as impurities after the synthesis, or introduced during the materials service, can affect various properties of alloys by changing the stability and mobility of crystalline defects[1–5]. One characteristic example is bcc refractory alloys based on group V (V, Nb, Ta) and VI (Mo, W) elements. Different from other alloys containing secondary precipitate phases, bcc refractory alloys are usually composed of a single bcc solid-solution phase, of which many properties are mainly managed by controlling the interactions of crystalline defects with solute elements, especially transition metal elements. For instance, the formation and growth of detrimental clusters, voids, and loops during high-energy neutron irradiation are strongly influenced by solute-vacancy interactions[6]. The strength and plasticity of bcc metals can be tailored by alloying with other transition metal elements due to a solute-induced effect on the dislocation mobility and twinnability[4,7–10]. Their grain boundaries (GB) can be further stabilized by solute segregations, thereby mitigating the driving force for grain growth at elevated temperatures[11].

Solute-defect interactions can be quantitatively characterized by the solute-defect binding energy. In a view of classical mechanics, the binding energy is usually correlated with the atomic volume or elastic moduli of solute elements, as the solute-defect binding usually relieves the strain energy penalty caused by the lattice mismatch between solute and matrix atoms[12–14]. Examples can be found in both non-transition (e.g. Al[13] and Mg[15]) and transition metal systems (e.g. Ni[16] and Co[17]). On the other hand, the atomic bond strengths and stress states at defects would also be affected by solute elements that induce variations in the local electron densities. This solute-induced variation in chemical bonding may also contribute to the solute-defect binding energies, so this variation is usually referred to as "*electronic contribution*" in the literature[17,18].

The solute-defect binding in bcc refractory metals seems to show strong dependences on the electronic features of solute elements, instead of the strain mismatch. A unique regularity — the solute-defect interaction becomes more attractive when the solute element has more valence electrons — has been reported for the interactions between transition metal elements and various types of crystalline defects in W/Mo alloys in different dimensions, including vacancies[19], dislocations[9,20], and GBs[21]. Moreover, the solute-defect interactions in V and Nb alloys show a distinct trend than those in W alloys: the solute-defect binding energy has a parabolic relationship with the increasing number of the valence electrons of the solute elements[22,23]. These results indicate that the solute-defect interactions in bcc refractory metals may have a universal electronic origin. The corresponding mechanisms remain elusive though there are some explanations for individual cases[19,20].

Using first-principles calculations based on density functional theory (DFT), herein we show that the binding behavior between transition-metal substitutional solute elements and various types of crystalline defects (0D, 1D, and 2D) in non-magnetic bcc refractory metals universally originates from the local variation in the electronic structure of the matrix atom at the substitutional site near the defect which is largely determined by two electronic factors. One factor is the variation in the bimodality feature of the *d*-orbital local density of states (LDOS) of the matrix atom before substitution; the other is the change in the bond hybridization strength between the valance *sp*- and *d*-bands. Moreover, based on these two electronic factors, a linear regression model is proposed, which is capable to describe the solute-defect interaction energies in binary alloys of bcc refractory metals with transition-metal substitutional solutes. Remarkably, this linear relationship is independent of the type of defects and the location of atomistic sites occupied by solutes. The present findings provide a fundamental basis for rational bcc refractory alloy design where the observed universality of the solute-defect interaction can quantitatively guide the proper selection of solute elements to control the energetic stabilities and mobility of the crystalline defects in alloys with targeted properties.

## Results

**Solute-dislocation interaction energy.** Fig. 1 shows the calculated interaction energy (i.e. binding energy) $E_{\mathrm{int}}$ between the $\frac{1}{2}\langle 111\rangle$ screw dislocation core and five types of transition metal substitutional solutes in bcc W, namely Ta, Re, Os, Ir and Pt. The interaction energies are calculated under two conditions: relaxing and fixing atomic positions during the total energy calculations of the solute-doped dislocation structures. It is worth to mention that the dislocation structure is fully relaxed to reach its equilibrium state in pure W and subsequently used for solute substitution. Therefore, the difference between the relaxed ($E_{int}^{relax}$) and fixed-lattice interaction energies ($E_{int}^{fix}$) gives the energy gained by the relaxation of the W lattice upon the solute substitution. As shown in Fig. 1, both the relaxed and fixed-lattice interaction energies show a strong correlation with the group number of the solute elements. The energies become more positive when the solute has more *d* electrons than W, while they become negative for the solute with fewer *d* electrons. More interestingly, the relative difference between $E_{int}^{relax}$ and $E_{int}^{fix}$ is small for all the solutes. These results indicate that the observed dependence of the interaction energies on the group number of solute element mainly originates from the local changes in the electronic structure and chemical bonding near the dislocation core rather than the effects of the lattice relaxation upon the solute substitution.

**Bimodality of the *d*-band LDOS of the atoms near dislocation core.** Owing to the localized characteristics of $d$ orbitals, the LDOS of transition metals can display considerable shape features that are characteristic of the given crystal structure[24,25]. Using W as an example, Fig. 2a shows that the bcc structure results in a bimodal *d*-band LDOS (solid-blue line) with a pseudo-band gap in the middle of the *d*-band, while the LDOS of close-packed structures (i.e. fcc/hcp) has a unimodal shape (solid-orange line). Interestingly, it is found that the LDOS of the W atom surrounding the screw dislocation core also has a less bimodal shape compared to that of perfect bcc (dashed-blue line), as a consequence of the change in local atomistic structures. Similar variation in LDOS is also observed for the $\frac{1}{2}\langle 111\rangle$ screw dislocation in Nb and Mo[26]. The bimodality distinction of LDOS was found previously to be essential for differentiating the energetic stabilities between the bulk phases with bcc and close-packed structures in transition metal systems[24,25,27]. When *d*-band is about half-filled, the Fermi level ($E_f$) is located close to the minimum of the pseudo-band gap in the LDOS of bcc structure, as shown in Fig. 2a. Qualitatively speaking, the LDOS of bcc structure has more occupied states far below $E_f$ and less occupied states close to $E_f$ compared to that of fcc structure when the $d$ band is about half-filled[25]. This leads to a lower electronic band energy, which makes bcc structure more stable compared to the close-packed structure[25].

Interestingly, it is found that solute substitutions do not change the bimodality features of LDOS for the dislocation core and the bcc bulk site, showing characteristics of the so-called *"canonical" d*-band[24,25,28]. Figs. 2b, 2c and 2d show the LDOS of atoms at a dislocation core site and a bulk bcc site far away from the core when these sites are occupied by Ta, Os or Pt instead of W, respectively. As shown in Figs. 2b-d, the solute atom at the core site still has a less bimodal LDOS compared with its counterpart at the bulk site. However, the filling fraction of the local *d*-band of the solute atom is changed as it has a different number of valence electrons than W. As shown in Figs. 2b and 2c, for the solutes with more *d* electrons, the position of the Fermi level ($E_f$) on LDOS shifts away from the middle pseudo-band gap towards the right band edge. According to bond-order potential theory, a structure with less bimodal DOS can usually be stabilized when the filling fraction is towards to the band edges, while a more bimodal DOS is favored for a half-filled band[24,25,27]. Therefore, compared to placing W atoms at the core site, the system may benefit from a stabilization contribution from the band energy when the core site is occupied by the solute atom with more *d* electrons than W. Correspondingly, there is an attractive interaction tendency between the dislocation core and these solute elements as shown in Fig. 1. A similar solute-induced stabilization mechanism has also been demonstrated on the $\{112\bar{1}\}$ twin boundary of hcp Re[29]. On the other hand, compared to that of the W atom, $E_f$ shifts to a position even closer to the minimum of the pseudo-band gap of the LDOS of the Ta solute as shown in Fig. 2d. Since the difference in the number of the occupied state close to $E_f$ between the core and bulk LDOS may be maximized at the minimum of the pseudo-band gap, Ta atom should be less preferred by the core site than W atom by considering occupied states close to and far below the $E_f$. This consequently yields a negative interaction energy as shown Fig. 1.

**Electronic attributes of solute-defect interactions.** The results in Figs. 1 and 2 reveal a qualitative correlation between the *d*-band bimodality and the solute-dislocation interaction in the binary alloys of bcc W and transition metal solutes. To further explore this correlation, we investigate the local electronic structures of atoms near several 0D, 1D and 2D defects in pure W, including mono-vacancy, <100>-dumbbell, <111>-dumbbell, $\frac{1}{2}\langle 111\rangle$ screw dislocation, $\Sigma3\ (11\bar{2})$ twin boundary (TB), Σ3(111), Σ5(310) and Σ5(210) GBs. To quantify the bimodality of the DFT-calculated LDOS, Hartigan's dip test was performed[30,31] (See *Method* section for details). A completed unimodal LDOS corresponds to a test statistic of 0, while a more bimodal LDOS has a larger value of test statistic[30,31]. We then use a parameter, $\Delta dip$, to quantify the change in the bimodality of the LDOS of the atoms near the defect relative to a reference atom that is far away from the defect. Correspondingly, $\Delta dip$ is defined as the difference in the dip test statistic between the LDOSs of the reference and defect atoms (i.e. $\Delta dip = dip(\text{reference}) - dip(\text{defect})$). Therefore, W atom at a site with a more positive $\Delta dip$ will have a less bimodal LDOS compared to the atom at the reference site. The calculation results are summarized in Table S2. It is found that all the investigated defect structures in the pure W configuration share a similar trend in their local electronic structures: W atoms near the defect center generally exhibit a less bimodal LDOS compared to those far away. Here we show the results of the mono-vacancy, $\frac{1}{2}\langle 111\rangle$ screw dislocation and $\Sigma3(11\bar{2})$ TB as three examples in Figs. 3a-c, respectively, where $\Delta dip$ of each defect atom is plotted with respect to its relative distance to the defect center.

Furthermore, for the W atoms where the $\Delta dip$ calculations are performed, we also calculate the corresponding fixed-lattice solute-defect interaction energies ($E_{int}^{fix}$) when these defect atoms are substituted by the Pt, Re and Ta solutes, respectively. The results are summarized in Table S2. In addition, the relaxed solute-defect interaction energies ($E_{int}^{relax}$) are also calculated for several defect sites to investigate the effect of lattice relaxation on the interaction energy. As shown in Table S5, like the solute-dislocation interactions, the difference between $E_{int}^{fix}$ and $E_{int}^{relax}$ are also small for other defect structures in W. By comparing the calculated $\Delta dip$ with $E_{int}^{fix}$, we notice a very interesting phenomenon that the variations in $E_{int}^{fix}$ of the Re and Pt solutes are strongly correlated with the variations in the bimodality of the LDOS for the W atoms being substituted at different defect sites. For example, as shown by the dashed lines in Figs. 3a-3c, the defect site with a higher $\Delta dip$ generally have a more attractive interaction with the solutes (higher $E_{int}^{fix}$). This correlation is consistent with the analyses in Fig. 2, since a more positive $\Delta dip$ corresponds to a less bimodal LDOS feature for W atom at that site. If we assume the solute substitutions do not change the bimodality features of LDOS as shown in Fig. 2, a less bimodal LDOS indicates that this atomic site prefers to be occupied by the solute atoms with more *d* electrons than W as $E_f$ will be a position closer to the edge of their *d*-band. Additionally, the correlation between $\Delta dip$ and $E_{int}^{fix}$ is found to be also valid for the Re and Pt solutes interacting with the defects in transition or unstable states, such as the generalized stacking faults (GSF) shown in Fig. S5 and S6 of the Supplemental

Material.

Moreover, if we plot all the calculated $E_{int}^{fix}$ together with respect to the corresponding $\Delta dip$ parameter, an approximately linear relationship with almost zero intercepts can be revealed between $E_{int}^{fix}$ and $\Delta dip$ for both Re- and Pt-substitutional solutes, as shown in Figs. S7a and S7b, respectively. These results indicate that the filling energy of the $d$-band associated with the bimodality variation indeed has significant contribution to the solute-defect interaction energy. The contribution can be quantitatively described by the $\Delta dip$ parameter. However, it should be pointed out that apparent deviations from the linear relationship are also observed for a few individual defect sites. Moreover, in contrast to the W-Re and W-Pt systems, it is found that the correlation between $E_{int}^{fix}$ and $\Delta dip$ in the W-Ta system becomes qualitative. For example, as shown in Fig. 3d, the Ta solute interacts in a repulsive way with the W $\Sigma 3(11\bar{2})$ TB generally, which yields a negative correlation between $E_{int}^{fix}$ and $\Delta dip$ ($\Delta dip > 0 \rightarrow E_{int}^{fix} < 0$), consistent with the analyses in Fig. 2d. However, quantitative discrepancies can be seen for several individual sites near the defects. As shown in Fig. 3d, the interaction energy of site 1 is about five times smaller than that of site 2, while the values of $\Delta dip$ between these two sites only vary about 20%. In addition, sites 4 and 5 in $\Sigma 3(11\bar{2})$ TB shown in Fig. 3d have nearly zero values of $\Delta dip$ and notable values of $E_{int}^{fix}$ in contrast. Therefore, from the above analyses, on the one hand, we may conclude that the variation in $d$-band bimodality ($\Delta dip$) is an electronic factor that plays an essential role in characterizing the solute-defect interaction in the binary alloys of bcc W and transition metal solutes. On the other hand, there are other underlying mechanisms that contribute to the solute-defect interaction energies, which cannot be described by the $\Delta dip$ term.

One possibility could be the energy contributions from the valence *sp*-band. Besides the importance of the *d*-band, the valence *sp*-band were found to be crucial for accurate description of many fundamental physical properties of transition metal elements, such as cohesive energy[32], equilibrium atomic volume[33,34] and bulk modulus[33,34]. Due to the covalent feature of the *d*-band, the valence *sp*-band can be strongly hybridized with and thus strongly influenced by the valence *d*-band. Within a tight-binding framework[32,33,35–39], the strength of the *sp*-*d* hybridization ($E_{sp}$) of an atom in transition metal alloys can be correlated with a function of (*i*) the interatomic distances between the atom and its neighboring atoms ($d_{ij}$) and (*ii*) the spatial extents of the d-orbitals of the atom and its neighboring atoms ($r_{d_i}$&$r_{d_j}$), which is $E_{sp} \propto \sum_j r_{d_i}^{\frac{3}{2}} r_{d_j}^{\frac{3}{2}} / d_{ij}^5$ (See *Section 7* in Supplemental Material for details). This suggests the strength of the *sp*-*d* hybridization in a defect structure should vary with each individual atom since $d_{ij}$ of the atom at each defect site can be different and the $r_{d_i}$ of the solute element can differ from that of the neighboring matrix element, as it is an intrinsic-element property. Therefore, the effect of the *sp*-*d* hybridization may not be ignored for determining solute-defect interactions in the bcc refractory alloys.

**Universal correlation between electronic factors and solute-defect interaction energy.** Based on the discussion above, we propose a linear regression model that considers the solute-defect interaction energy ($E_{int}^{fix}$) in two parts as shown in Eq. 1,

$$E_{int}^{fix} = \Delta E_d + \Delta E_{sp} \approx a_1 \Delta dip + a_2 x_{sp} \qquad \text{Eq. 1}$$

Here, $\Delta E_d$ represents the energy contribution due to the $d$-band filling. Based on the results of Fig. S7, this part of energy may linearly correlate with the changes in the bimodality of the $d$-band through the $\Delta dip$ term and a fitting coefficient, $a_1$. The second part in Eq. 1, $\Delta E_{sp}$, represents the energy contribution related to the *sp*-*d* hybridization. We propose $\Delta E_{sp}$ can also be estimated through a fitting coefficient, $a_2$, and a variable, $x_{sp}$, that describe the local environment of the defect site related to the *sp*-*d* hybridization. Like the $\Delta dip$ term, it is expected that $x_{sp}$ can also be obtained from the DFT calculations of the defects in pure metals. Then, the alloying effects can be reflected by the fitting coefficient $a_1$ and $a_2$.

In the present work, $x_{sp}$ of an atom near the defect in pure metals is proposed to be,

$$x_{sp} = 1 - \frac{\left(V_{vor}^{def}\right)^{-\frac{5}{3}} / \epsilon_{sp}^{def}}{\left(V_{vor}^{ref}\right)^{-\frac{5}{3}} / \epsilon_{sp}^{ref}} \qquad \text{Eq. 2}$$

where $V_{vor}^{def}/V_{vor}^{ref}$ is the Voronoi volume of the atom at the defect and reference site, respectively, and $\epsilon_{sp}^{def}\,/\epsilon_{sp}^{ref}$ is the center of the occupied *sp*-band projected on the atom at the defect and the reference site, respectively. The reference site is same as the one used for the calculation of $\Delta dip$ and $E_{int}^{fix}$. The $\epsilon_{sp}^{def}$ term is calculated as,

$$\epsilon_{sp}^{def} = \int_{-\infty}^{0} E\rho_{sp}^{def}(E)dE \,/\, \int_{-\infty}^{0} \rho_{sp}^{def}(E)dE \qquad \text{Eq. 3}$$

where $\rho_{sp}^{def}(E)$ is the projected LDOS of the *sp*-band on the atom at the defect site and the Fermi energy $E_f$ is set to zero. $\epsilon_{sp}^{ref}$ is calculated in the same way for the atom at the reference site. The Voronoi volume and LDOS of the *sp*-band are calculated from the relaxed atomic structures of pure matrix metals that contain defects. In Eq. 2,

Voronoi volume ($V_{vor}$) is used to describe the average changes in the interatomic distances ($d_{ij}$) of the atoms near the defect. The exponent value, $-\frac{5}{3}$, is obtained since $E_{sp}$ is proportional to $\left(d_{ij}\right)^{-5}$ (See *Section 7* in Supplemental Material for details). A benefit of using Voronoi volume instead of directly calculating $d_{ij}$ is to avoid arbitrary assignment of the cutoff distance for identifying neighboring atoms. In Eq. 2, we also include a scaling term, $1/\epsilon_{sp}$. Based on the definition (Eq. 3), $\epsilon_{sp}$ actually characterizes the average difference between the energy states of the *sp*-band and $E_f$. Generally speaking, the electrons with energies close to $E_f$ may be more sensitive to small perturbations. We thus propose that, for the *sp*-band closer to the $E_f$, its corresponding *sp-d* hybridization may have a stronger effect on the ground state energy. Therefore, the inverse of $\epsilon_{sp}$ is used to the scale the effects of *sp-d* hybridization on solute-defect interactions. It is noteworthy that the inclusion of the $\epsilon_{sp}$ term in Eq. 2 is necessary as we found that the Voronoi volume alone is inadequate to construct the $x_{sp}$ term to yield an accurate description of the solute-defect interaction energy. In addition, attempts have been made to adopt the electronic factor descriptors based on other common features of the electronic bands, such as the *d*-band center, zero-, first- and second-order moments of the *sp*- and *d*-band and so on. However, $\Delta dip$ and $x_{sp}$ are the two descriptors so far that yield the most accurate regression results.

Based on Eq. 1, we perform linear regressions to model the DFT-calculated $E_{int}^{fix}$ of the crystalline defects in the W-Ta, W-Re and W-Pt binary alloy systems, where W is the matrix element. $\Delta dip$ and $x_{sp}$ are treated as regression variables, which are obtained for each defect site from the corresponding DFT defect calculations in pure metals. The values of $\Delta dip$ and $x_{sp}$ of each defect site are listed in Table S2. $a_1$ and $a_2$ are fitting coefficients, which should vary for each alloy system to reflect the effects of the solute elements. The least-squares fitting method is employed to perform the regression. The regression parameters for each matrix-solute element pair are summarized in Table 1. As shown in Fig. 4, the solute-defect interaction energies ($E_{int}^{fix}$) predicted by the proposed regression model show good agreement with the results of DFT calculations for the W alloys with different transition metal solutes (i.e. Ta, Re and Pt). Good regression quality is also demonstrated by the close-to-one value of adjusted $R^2$ and a small value of the standard error, which is about 5% of the largest response value for each regression as listed in Table 1. In addition, it is found that the p-values of both variables (i.e. $\Delta dip$ and $x_{sp}$) are almost zero, significantly less than 0.05, which means both variables are statistically meaningful to explain the response variance.

Considering the closeness of the crystal and electronic structures between group V and VI bcc elements, one would naturally wonder whether Eq. 1 can also be universally applied to model the solute-defect interactions in the binary alloys of group V element and transition metal solutes. To explore the possible universality, we also perform DFT calculations to calculate the $\Delta dip$ and $x_{sp}$ of atoms in several 0D, 1D and 2D crystalline defects in pure Ta. As expected, it is found that Ta atoms near the defect center also generally have a less bimodal LDOS compared to those far away. For example, the *d*-orbital LDOS for a Ta atom exactly on the interface plane of the $\Sigma3(11\bar{2})$ TB are plotted in Fig. 5a, along with the LDOS of a Ta atom at the reference site that is far away from the interface. The latter LDOS indeed shows more bimodal characteristics with a deeper pseudo-band gap than the former LDOS.

The fixed-lattice solute-defect interaction energies ( $E_{int}^{fix}$ ) are also calculated correspondingly when Ta atoms are substituted by the Hf and Os solutes. Similar to the cases for W matrix, it is found that the lattice relaxation also has very minor effects on the solute-defect interaction energy (see Table S4 in Supplementary Materials for details). The obtained values of $\Delta dip$, $x_{sp}$ and $E_{int}^{fix}$ of each defect site are listed in Table S3. Then, linear regressions based on Eq. 1 are performed to model the DFT-calculated $E_{int}^{fix}$. Parity plots of the regression results are shown in Figs. 5b and 5c for Ta-Hf and Ta-Os systems, respectively. The regression coefficient and parameters are listed in Table 1. As shown by both Fig. 5 and Table 1, the proposed regression model (Eq. 1) can be generally applied to quantitatively describe the solute-defect interactions in Ta-based alloys as well.

Additionally, it is worth to note that there are some apparent discrepancies between the model-predicted and DFT-calculated $E_{int}^{fix}$ for a few individual defect sites. For example, as shown in Fig. 4c, Figs. 5b and 5c, the solute-interactions of the <100>-dumbbell defect seems to be poorly described in the W-Pt, Ta-Hf and Ta-Os systems, although the variation trend is qualitatively reflected. A possible reason could be the approximation on using the Voronoi volume to averagely represent the interatomic distance is not appropriate here, since this average may have large uncertainties in terms of the standard deviation when defect-induced lattice distortions become large and anisotropic. Another possible reason is that the correlation between the interaction energy and electronic factors could become non-linear for the atomic site with large electronic/lattice perturbations. A future investigation is deserved to discover more appropriate electronic descriptors and establish more accurate regression models.

**Disparities and similarities between group V and VI bcc matrix elements.** By further scrutinizing the fitting coefficients in Table 1, it is found that the $\Delta dip$ term yields a positive contribution to interaction energy for the solute with more $d$ electrons than W (e.g. Re and Pt), while it yields a negative contribution for the solute with fewer $d$ electrons (e.g. Ta), which is consistent with our analysis in Fig. 2. However, the contribution becomes positive in the Ta alloys for the solute with fewer $d$ electrons than Ta (e.g. Hf), while becoming negative for the solute with slightly more $d$ electrons (e.g. Os). This is because that the relative position of $E_f$ on the LDOS of the $d$-band is intrinsically different between Ta and W when they serve as the matrix element. As shown in Fig. 5a, $E_f$ of the Ta matrix is located at a position close to the lower energy of the bcc pseudo-band gap, but $E_f$ of the W matrix shifts to the right at a position within the pseudo-band gap due to a higher filling fraction (Fig. 2a). Therefore, on the one hand, when alloying Ta with transition metal elements with fewer $d$ electrons, such as Hf, the position of $E_f$ on the local $d$-band of the solute atom would further shift away from the pseudo-band gap to the left band edge compared to that of Ta matrix atom, leading to a positive contribution in terms of the $\Delta dip$ parameter. On the other hand, negative contributions can be expected for the solute with slightly more $d$ electrons (e.g. Os) due to a relocation of $E_f$ into the pseudo-band gap. Moreover, one can imagine that a solute-induced stabilization effect similar to W could also be triggered if $E_f$ continues moving across the pseudo-band gap to the right band edge by alloying Ta with the solute element having even more $d$ electrons (e.g. Au). As a result, the energy contributions of the $\Delta dip$ term in the alloys based on group V elements may have an overall parabolic relationship with the number of $d$ electrons of solute, which may be reflected in some cases of the solute-defect interactions (e.g. Fig. S10 and Ref. 22 and 23).

In addition, the coefficient of the $x_{sp}$ term also has a dependence on the number of $d$ electrons of the solute element, which is consistent in both Ta- and W-based alloys. As shown in Table 1, $a_2$ generally has a positive sign if the solute element has less d electrons than the matrix element (e.g. W-Ta and Ta-Hf), while it yields a negative sign if the difference in the number of $d$ electrons is reversed. This correlation can be understood in terms of the difference in the spatial extent of d-orbital between solute and matrix elements. Details are explained in *section 9* of Supplementary Materials. As indicated by the above analyses, the regression coefficients of the $\Delta dip$ and $x_{sp}$ term both have strong dependences on the $d$-orbital features of the solute elements. A summary of this electronic dependence is shown in Fig. 6, which may provide a useful guidance for the proper selection of solute elements to control the energetic stabilities and mobility of the crystalline defects in bcc refractory alloys.

**Discussion**

Based on first-principles calculations we have shown that the solute-defect interactions in the binary alloys of bcc refractory metals with transition metal solutes are universally originated from two electronic factors: ($i$) the change in the bimodality of the local $d$-band ($\Delta dip$) and ($ii$) the variations in the hybridization features of the local valence $sp$-band ($x_{sp}$) of the atom near the defect. For a specific solute element in a particular refractory metal, the interaction energy ($E_{int}^{fix}$) can be well assessed by a linear regression model solely based on the two electronic factors (i.e. $\Delta dip$ and $x_{sp}$). All these values ($E_{int}^{fix}$, $\Delta dip$ and $x_{sp}$) are calculated using the fixed atomistic structures of defects that are already optimized by DFT calculations in pure matrix metals. The observed correlation may provide a new way to efficiently predict the solute-defect interaction energy of any atomic site near the defect. For example, as shown in Fig. S11, with the linear relationship established based on the calculations for a few defects with relatively simple atomistic structures, one can estimate the interaction energy for complex defect structures with reasonably small uncertainty. In another word, instead of running many case-by-case DFT calculations for different substitutional configurations, only one DFT calculation for the defect in pure metal configuration is required for obtaining the $\Delta dip$ and $x_{sp}$ parameters. This could significantly reduce the computational costs for defects in highly complex geometry.

Moreover, since variations in both the bimodality of LDOS and the *sp-d* orbital hybridization are governed by the local atomic structure of the defects, in the future, it is possible to quantitatively connect the $\Delta dip$ and $x_{sp}$ parameters of an atom at a defect site with its local structural features, such as bond-orientation parameter[40], coordination number, radial distribution function and so on. Such connection can be achieved by applying machine learning methods[41,42]. As a result, one would be able to estimate the electronic factors (i.e. $\Delta dip$ and $x_{sp}$) of each atomic site in a defect geometry optimized from the atomistic simulations based on empirical interatomic potentials rather than first-principles calculations. Then, with the established correlation between the solute-defect interaction and the electronic factors, the variations in the energetic properties of the various types of crystalline defects due to transition-metal solutes can be efficiently predicted. This will be particularly useful for complex defect configurations critical to their kinetics, such as dislocation kinks/jogs and disconnections in grain boundaries[43,44], which are otherwise hard to directly study using first-principles calculations.

In summary, our findings establish a general and quantitative correlation between electronic structure factors and energetic stabilities of crystalline defects containing substitutional solute atoms in bcc refractory alloys. It is developed starting from the classical theories of bulk phase stability based on electronic structures and extended to explain the energetic stabilities of the local structural units at the atomistic level[29]. The characteristics of this correlation are independent of the substitutional solute elements, the defect types and the locations of substitutional sites. It can be applied to efficiently map solute-defect interactions by investigating the electronic and atomic structures of defects in pure metals. It also provides possibilities to predict solute-defect interactions based only on the atomic structures of defects in pure metals if further studies on the correlations between local electronic and atomistic factors are carried out. This correlation can serve as a quantitative guideline for the transition metal alloy design with targeted properties by controlling the effects of solute-defect interactions on defect stability and mobility. From a broader perspective, the results from this study provide a robust example and a key step to construct advanced theories to describe the quantitative connections between the chemical bonding characteristics at the electronic level and the macroscopic materials properties[45–47]. In addition, the observed electronic factors have the potentials to serve as useful descriptors in data-centric science for materials innovation based on machine learning and artificial intelligence[48].

## Method

**First-principles calculations:** First-principles calculations in the present work were carried out using the projector augmented wave method (PAW)[49] and the exchange-correlation functional depicted by the general gradient approximation from Perdew, Burke, and Ernzerhof (GGA-PBE)[50], as implemented in the Vienna ab-initio simulation package (VASP)[51]. The energy cutoff of the plane-wave basis was 400 eV. Brillouin zone integration was performed using a first-order Methfessel-Paxton smearing of 0.2 eV[52]. The grid of the k-point mesh in the first Brillouin zone is set according to the size and geometry of the simulation supercells (See *section 2* in Supplementary Material for details). The convergence criterion was set as $10^{-7}$ eV/atom for the structure relaxation and $10^{-8}$ eV/atom for the static calculations. The electronic configurations of the pseudopotentials used for the present first-principles calculations are summarized in Table S1. As shown in Table S1, the semi-core *5p* electrons are treated as valence electrons for the calculations of Hf, Ta and W. However, it is found that the LDOS of the *5p*-band localizes at very low energy states far away from the Fermi level and has a very large energy gap with the *5d*-, *6s*- and *6p*-bands. We thus assume that the *5p* electrons are basically inner-core electrons that have very limited contributions to electronic bonding. Therefore, the LDOS of the *5p*-band is not included in the band analysis based on Eq. 3.

First-principles calculations are performed in three steps to model the local electronic factors of the crystalline defects in bcc Ta and W, and their interactions with substitutional solute atoms. In the first step, relaxation calculations are performed to obtain the optimized atomistic structures of crystalline defects in the pure metal matrix. In each relaxation calculation, the atoms and geometry of the simulation supercells are fully relaxed according to the Hellmann-Feynman forces, except calculations for the $\frac{1}{2}\langle 111\rangle$screw dislocation and the GSF defects due to their unique atomistic geometries. The relaxation of the $\frac{1}{2}\langle 111\rangle$ screw dislocation is performed using the flexible boundary condition method[53,54]. The relaxation scheme is described in details in the previous publication[4,9]. In the calculations of the GSF defects, the atoms are only allowed to relax along the direction perpendicular to the fault plane. In the second step, static calculations are performed based on the relaxed defect structures to obtain the projected local electronic density of state (LDOS) on each atom in the supercells. Then, the $\Delta dip$ and $x_{sp}$ parameters of each atomic site of interest are obtained from the DFT-calculated LDOSs and atomistic structures. In the third step, solute atoms are introduced to substitute the individual solvent atoms with different separation distances to the defect center to investigate the solute-defect interactions. The relaxed defect structures in pure metals are used for solute substitution. After substitution, the interaction energies are then calculated under two different conditions: fixing and relaxing atomic positions during the total energy calculations of the solute-doped defect structures. The difference between the relaxed ($E_{int}^{relax}$) and fixed-lattice interaction energies ($E_{int}^{fix}$) gives the energy change due to the relaxation of the defect lattice upon the solute substitution. The fixed-lattice interaction energies are calculated for all solute-defect interactions considered in the present work, while the relaxed interaction energies are only calculated for a few defect sites in order to evaluate whether the lattice relaxation has a significant contribution to the solute-defect interaction energies. A detailed comparison between the calculated $E_{int}^{fix}$ and $E_{int}^{relax}$ is described in *section 4* in Supplementary Material.

**Hartigan's dip test:** The Hartigan's dip test is a statistical method proposed by Hartigan and Hartigan[31], which measures the deviant of an empirical distribution function from unimodality. The test takes a sample from an empirical distribution function as inputs, and returns a positive dip test statistic. For a given sample size, the most favorable unimodal distribution corresponds to a statistic of 0, while a more significant bimodal distribution is evidenced by a larger statistic. In the present work, to perform the Hartigan's dip test, the LDOS from first-principles calculations was normalized with respect to its total number of density of states. The sample for the dip test was then drawn randomly from the normalized LDOS with a size of 500 data points (Each LDOS in the present work was set to have 301 energy intervals in first-principles calculations). We have drawn 8000 samples for each LDOS, and the dip test statistic of each LDOS being used for comparison is taken as the average of the statistics from the 8000 samples. All the Hartigan's dip tests of bimodality of LDOS were performed using a MATLAB code by Mechler[55].

## Data availability

The data and code that support the findings of this study are available from the corresponding author (qiliang@umich.edu) on request.

## Acknowledgement

Y.J.H, C.M.Y and Q.L. acknowledge support by startup funding from the University of Michigan. This research was supported in part through computational resources and services provided by Advanced Research Computing at the University of Michigan, Ann Arbor. This research used resources of the National Energy Research Scientific Computing Center, a DOE Office of Science User Facility supported by the Office of Science of the U.S. Department of Energy under Contract No. DE-AC02-05CH11231. B.Z. and X.Q. acknowledge the startup fund of Texas A&M University and the advanced computing resources provided by Texas A&M High Performance Research Computing. Finally, we would like to thank Professor Dallas R. Trinkle in University of Illinois Urbana-Champaign for sharing his simulation codes on the flexible boundary condition method.

## Author Contributions Statements

Y.J.H., X.Q. and L.Q. conceived of the research strategy and designed the modeling procedures. Y.J.H., B.Z., and C.Y. performed the first-principles calculations. Y.J.H. and G.Z. performed the Hartigan's dip tests. Y.J.H., Z.K.L., X.Q. and L.Q. prepared the manuscript. L.Q. supervised the project. All authors discussed the results and contributed to the manuscript.

**Table 1. Regression coefficients and parameters of the linear regression model based on Eq. 1.**

| Alloy system | coefficient | | Adjusted $R^2$ * | Standard error | p-value** | |
|---|---|---|---|---|---|---|
| | $a_1$ | $a_2$ | | | $\Delta dip$ | $x_{sp}$ |
| W-Ta | -7.18 | 1.78 | 0.9252 | 0.044 | 1.5e-23 | 1.3e-31 |
| W-Re | 15.98 | -1.29 | 0.9603 | 0.039 | 1.6e-56 | 2.0e-30 |
| W-Pt | 61.80 | -1.00 | 0.9453 | 0.168 | 3.4e-53 | 1.0e-03 |
| Ta-Hf | 6.66 | 2.08 | 0.9347 | 0.043 | 2.5e-19 | 6.0e-35 |
| Ta-Os | -6.43 | -4.06 | 0.8848 | 0.107 | 4.6e-06 | 5.3e-29 |

* The adjusted $R^2$ is the coefficient of determination, which represents the proportion of the variance in the regression response that is predictable from the regression variables. **The p-value tests the null hypothesis for each regression variable. A small p-value (typically $\leqslant$ 0.05) indicates strong evidence to reject the null hypothesis, which means the changes in the variable value is strongly related to the changes in the fitting response value ($E_{int}^{fix}$).

**Figure 1. Solute-defect interaction energy ($E_{int}$) between transition metal elements and the $\frac{1}{2}\langle 111\rangle$ screw dislocation in bcc W.** The interaction energy, $E_{int}$, is defined as the difference between the total energies of the dislocation structure with a solute atom X occupying an atomic site far away from and at the dislocation core. The dislocation structure is initially fully relaxed to reach its equilibrium state in pure W, and sequentially used for solute substitution. A positive value of $E_{int}$ represents an attractive binding tendency. The interaction energies are calculated under two conditions: relaxing ($E_{int}^{relax}$) and fixing atomic positions ($E_{int}^{fix}$) during the total energy calculations after the solute atom added into the supercell. The values of $E_{int}^{relax}$ are taken from our recent publication[9].

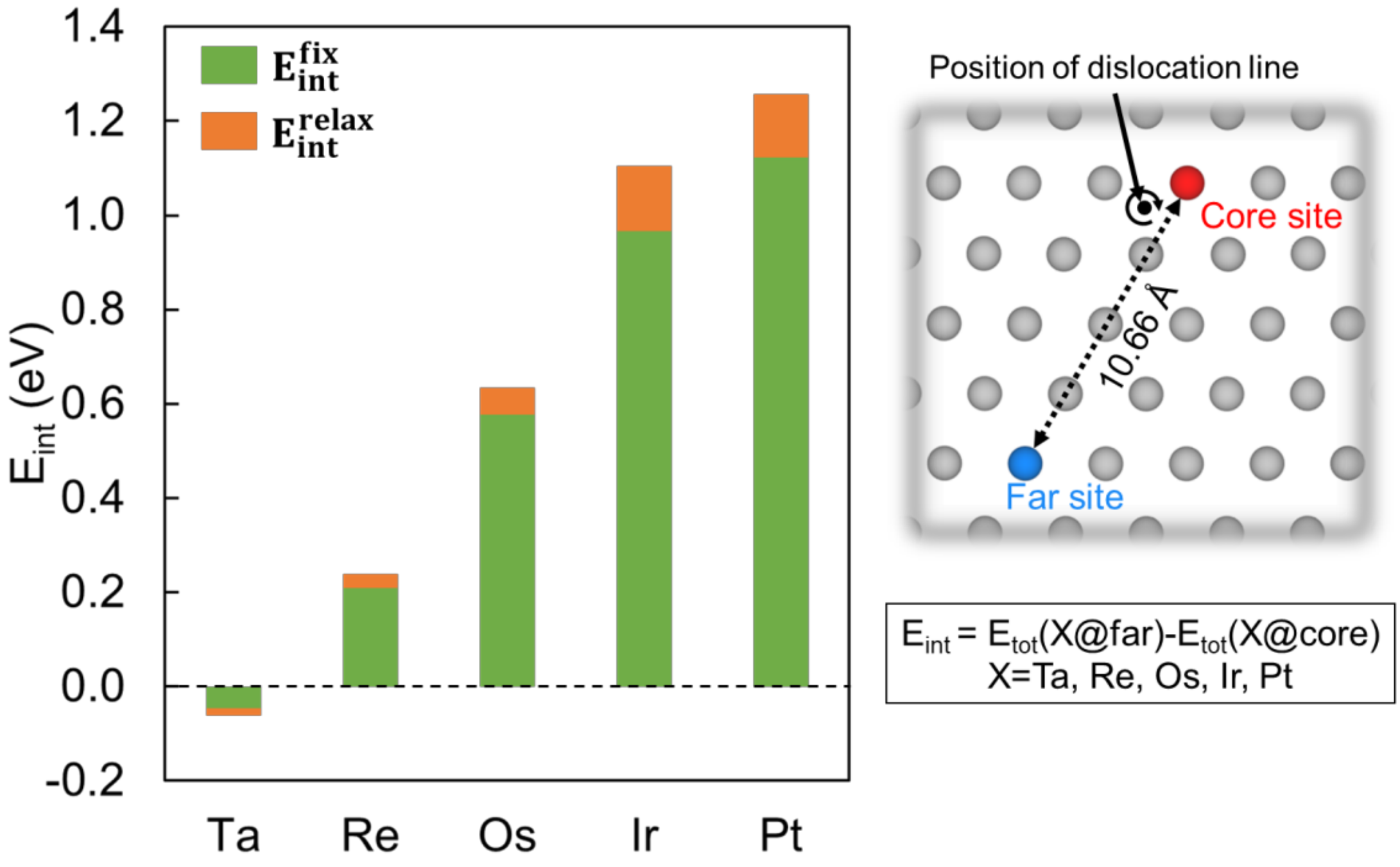

**Figure 2. Projected LDOSs of *d*-bands.** (a) LDOS of a W atom in perfect bcc lattice (solid-blue line), perfect fcc lattice (solid-orange line) and at the $\frac{1}{2}\langle 111\rangle$ dislocation core site (dashed-blue line) in pure W. (b)-(d) LDOS of an Pt, Os and Ta atom occupying the bcc site (solid-blue line) and the $\frac{1}{2}\langle 111\rangle$ dislocation core site (dashed-blue line) in the W matrix, respectively. The bcc bulk site and core site refer to the atomic sites marked in blue and red color in Fig. 1, respectively.

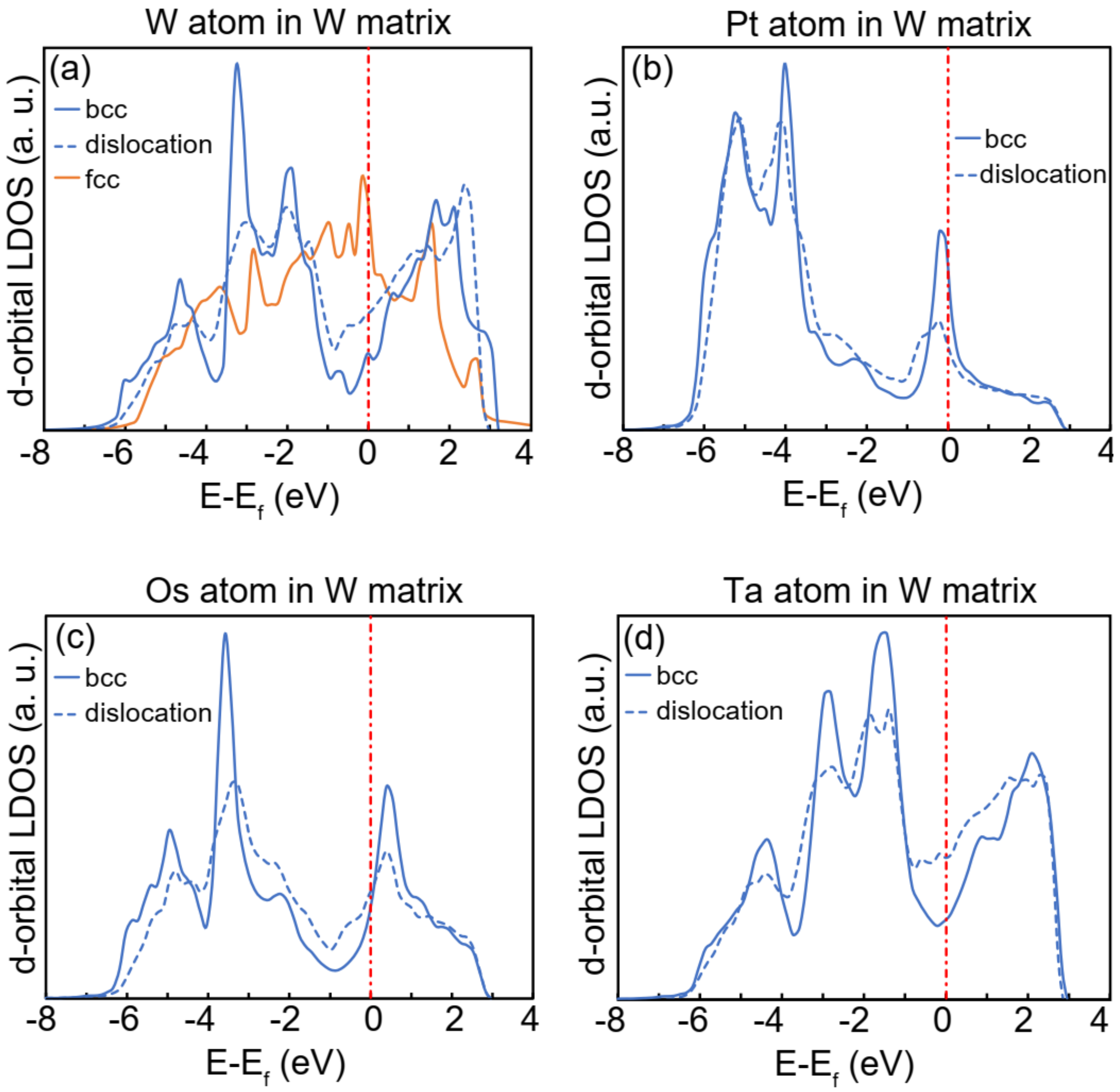

**Figure 3. Correlation between the defect-induced changes in bimodalities of LDOS ($\Delta dip$) in pure W and solute-defect interaction energies ($E_{int}^{fix}$) in W-based binary alloys.** (a) Vacancy; (b) $\frac{1}{2}\langle 111\rangle$ screw dislocation; (c) and (d) $\Sigma 3(11\bar{2})$ TB. $E_{int}^{fix}$ refers to the interaction energy that calculated based on the defect structures that are already fully relaxed in pure W and without further applying atomic position relaxations after solute substitution. The calculated $\Delta dip$ (■) of each atomic site of interest, and the corresponding solute-defect interaction energy when the site is occupied by Re (▲) Pt (●) and Ta (◆) are plotted with respect to the relative distance from the atomic site to the defect center. The positions of the atomic sites in the simulation cell are marked by numbers according to the pairing distance with the defect center and the investigated site. It should be noted that the axis value for the $\Delta dip$ term in Fig. 3d is plotted in reverse order (more positive values of $\Delta dip$ toward the downward direction). Completed configurations of the simulation cells can be found in Figs. S1-S3.

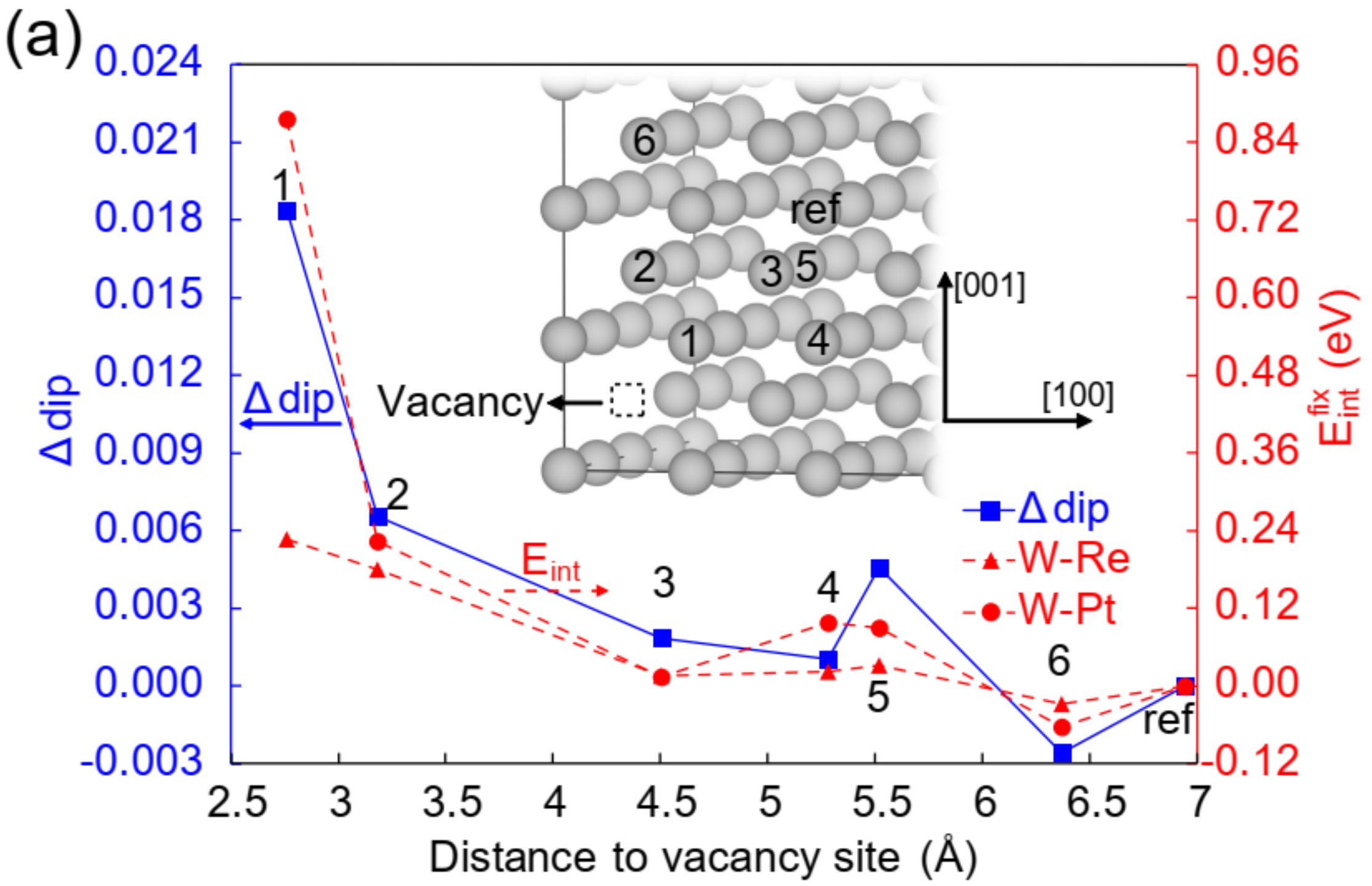

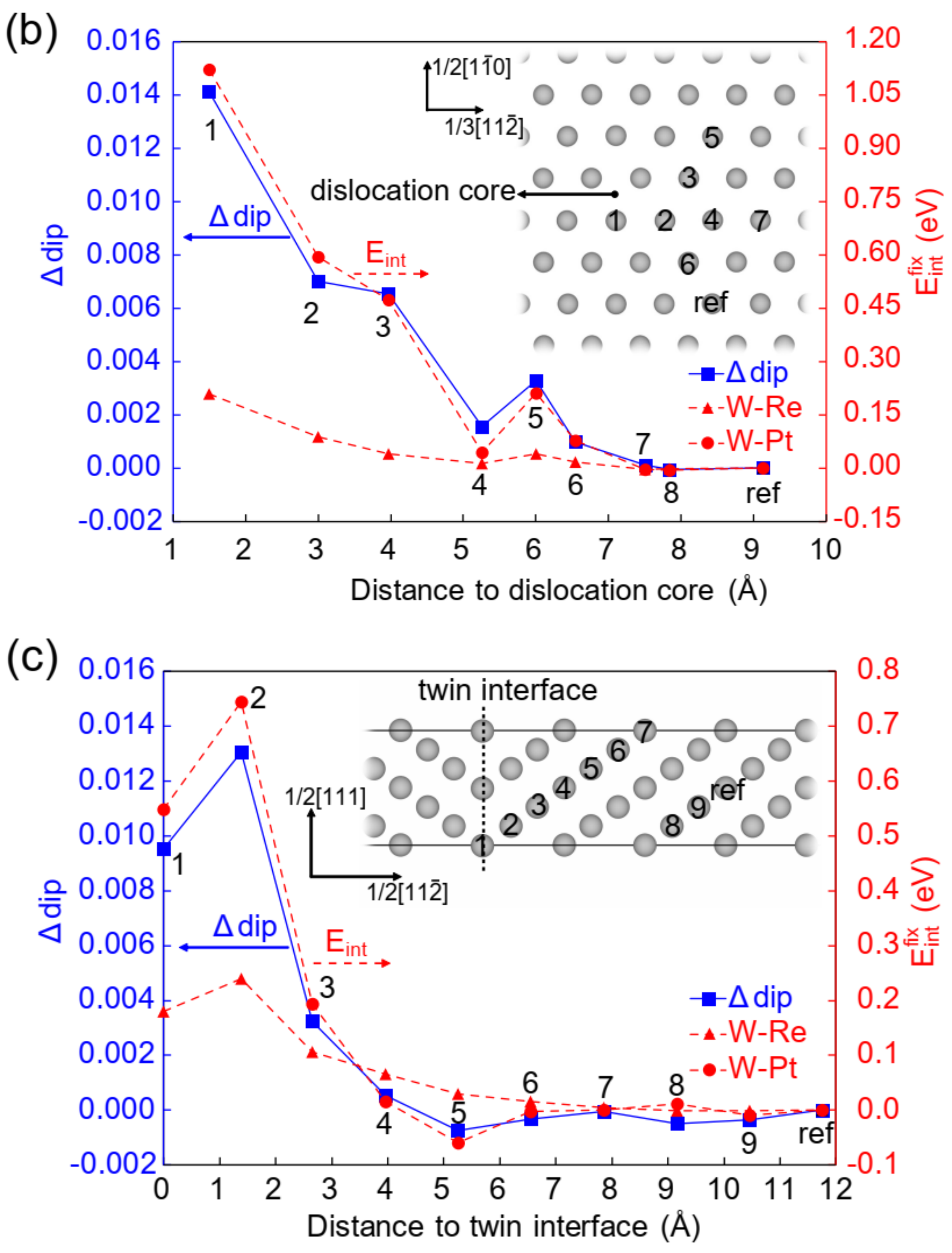

(b)
0.016
0.014
0.012
0.010
0.008
0.006
0.004
0.002
0.000
-0.002
Δdip
1.20
1.05
0.90
0.75
0.60
0.45
0.30
0.15
0.00
-0.15
$E_{int}^{fix}$ (eV)
1/2[1$\bar{1}$0]
1/3[11$\bar{2}$]
dislocation core
Δ dip
$E_{int}$
1
2
3
4
5
6
7
8
ref
Δdip
W-Re
W-Pt
1 2 3 4 5 6 7 8 9 10
Distance to dislocation core (Å)
(c)
0.016
0.014
0.012
0.010
0.008
0.006
0.004
0.002
0.000
-0.002
Δdip
0.8
0.7
0.6
0.5
0.4
0.3
0.2
0.1
0.0
-0.1
$E_{int}^{fix}$ (eV)
twin interface
1/2[111]
1/2[11$\bar{2}$]
Δ dip
$E_{int}$
1
2
3
4
5
6
7
8
9
ref
Δdip
W-Re
W-Pt
0 1 2 3 4 5 6 7 8 9 10 11 12
Distance to twin interface (Å)

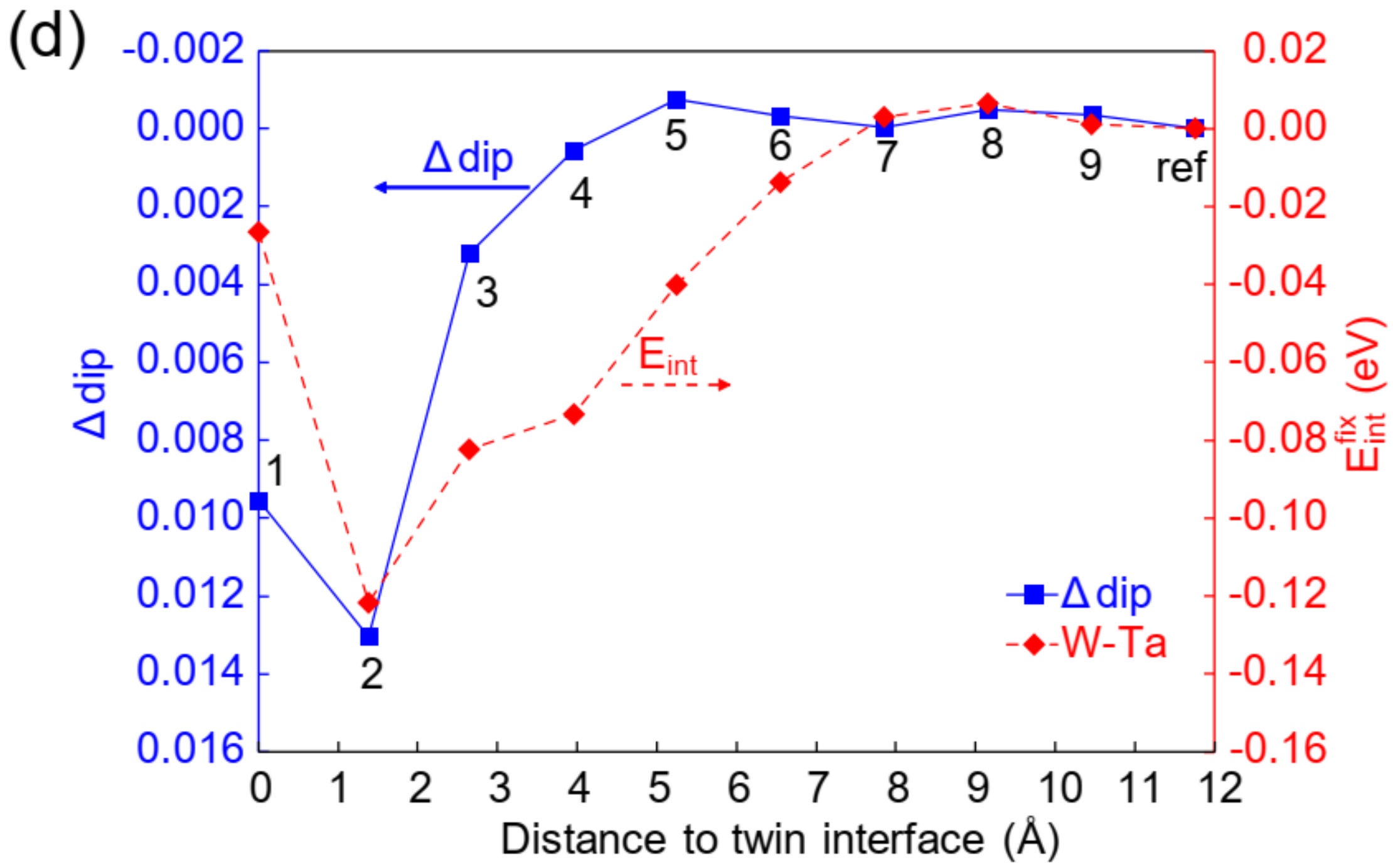
(d)
-0.002
0.000
0.002
0.004
0.006
0.008
0.010
0.012
0.014
0.016
Δdip
0.02
0.00
-0.02
-0.04
-0.06
-0.08
-0.10
-0.12
-0.14
-0.16
$E_{int}^{fix}$ (eV)
Δdip
$E_{int}$
1
2
3
4
5
6
7
8
9
ref
Δdip
W-Ta
0 1 2 3 4 5 6 7 8 9 10 11 12
Distance to twin interface (Å)

**Figure 4. Comparison between the $E_{int}^{fix}$ from DFT calculations and predicted from the regression model in the W-based binaries.** (a) W-Ta system; (b) W-Re system; (c) W-Pt system. The data of the point, line and planar defects are marked in circle, triangle and square symbols, respectively. The DFT-calculated $E_{int}^{fix}$ refers to the solute-defect interaction energies calculated based on fixed atomistic structures that are already fully relaxed in pure W.

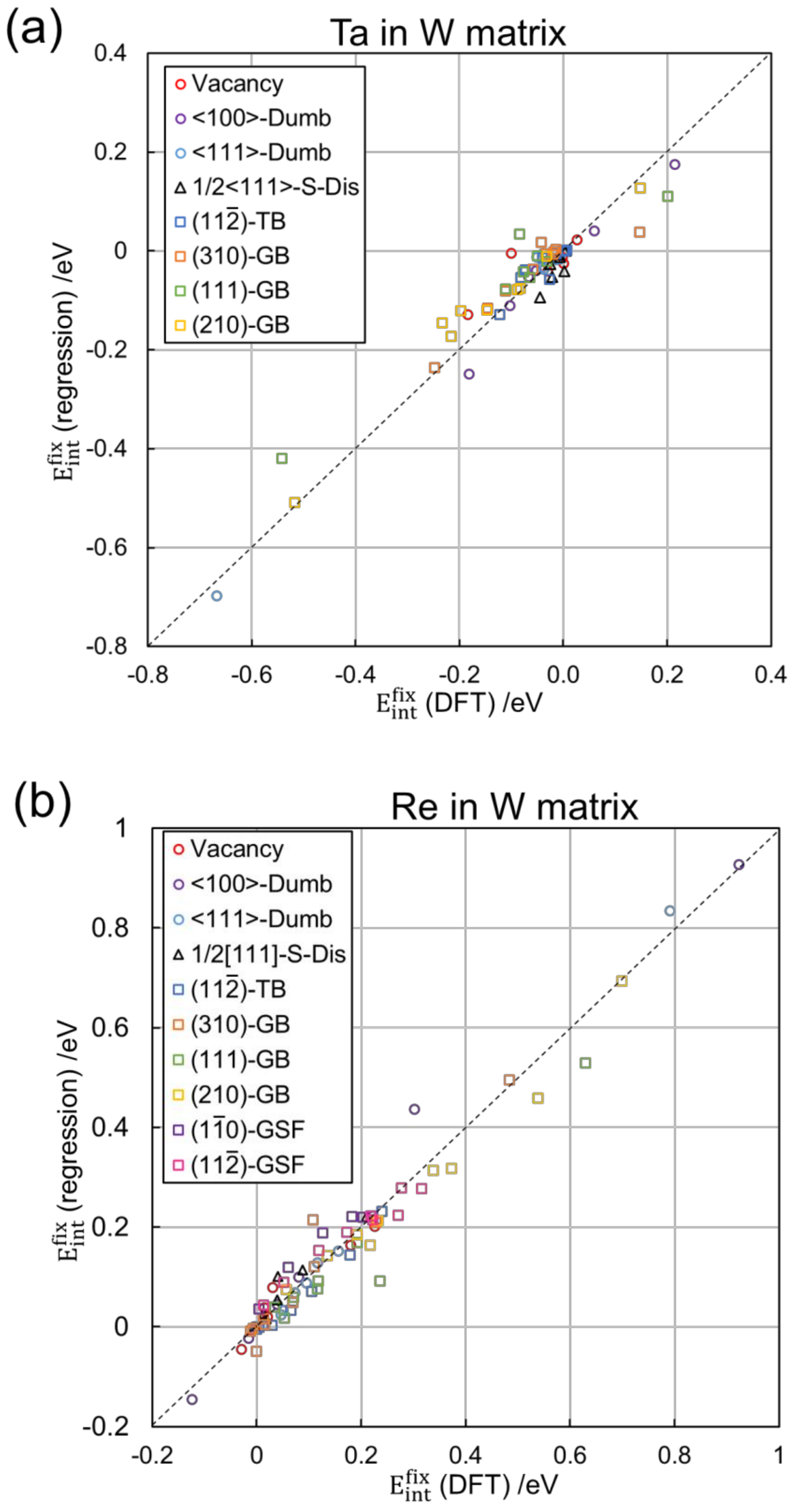

(c)

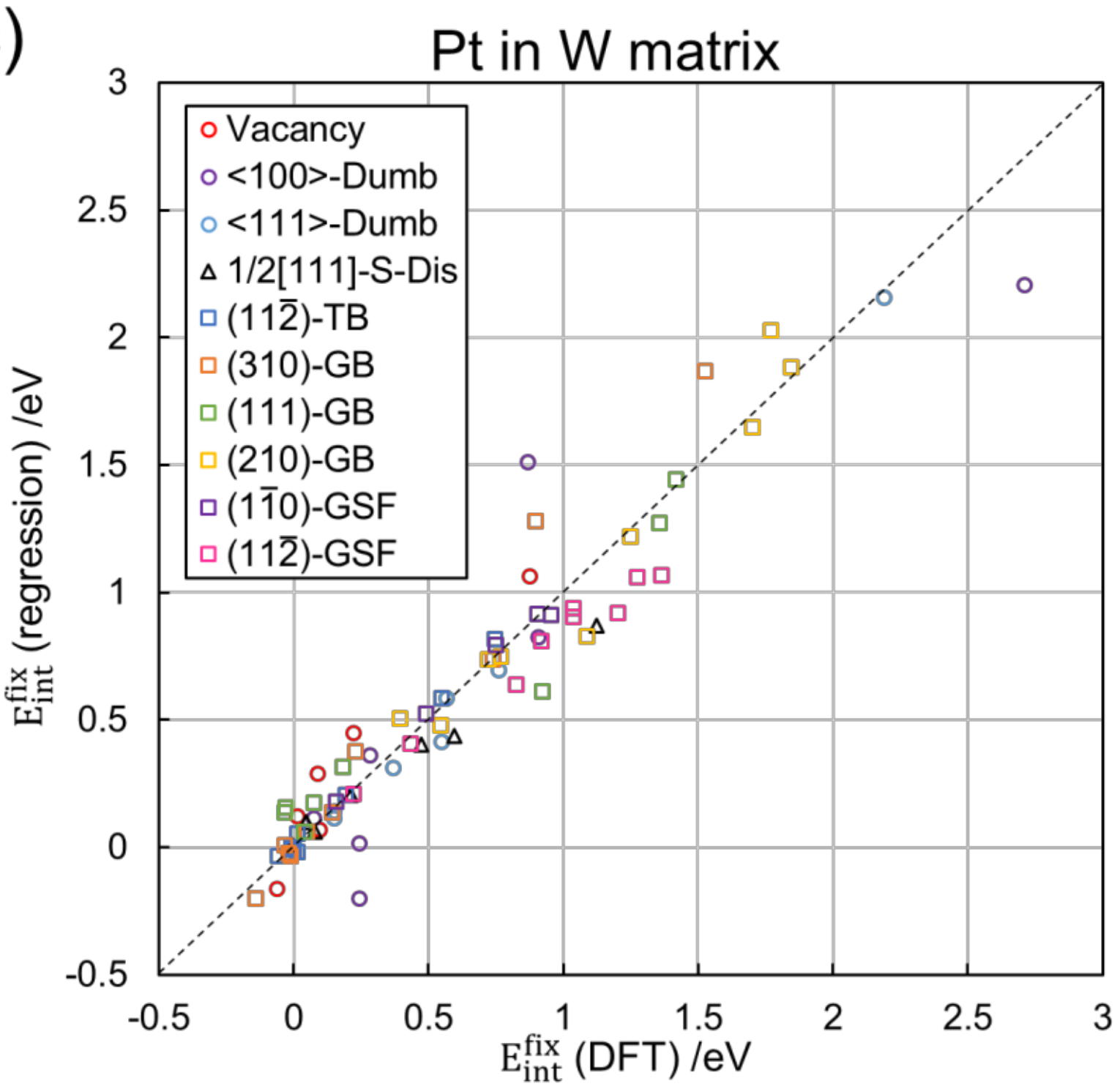

**Figure 5. *d*-band bimodality and solute-defect interaction energies in bcc Ta.** (a) Projected LDOSs of $d$ orbitals of a Ta atom on the interface of the $\Sigma3(11\bar{2})$ TB (dashed line) and in bulk lattice (solid line), respectively. The energy axis is scaled relative to the Fermi level. (b) and (c) DFT-calculated $E_{int}^{fix}$ in comparison with the predictions from the regression model in the cases of the Ta-Hf and Ta-Os systems, respectively. The DFT-calculated $E_{int}^{fix}$ refers to the solute-defect interaction energies calculated based on fixed atomistic structures that are already fully relaxed in pure Ta.

(a)

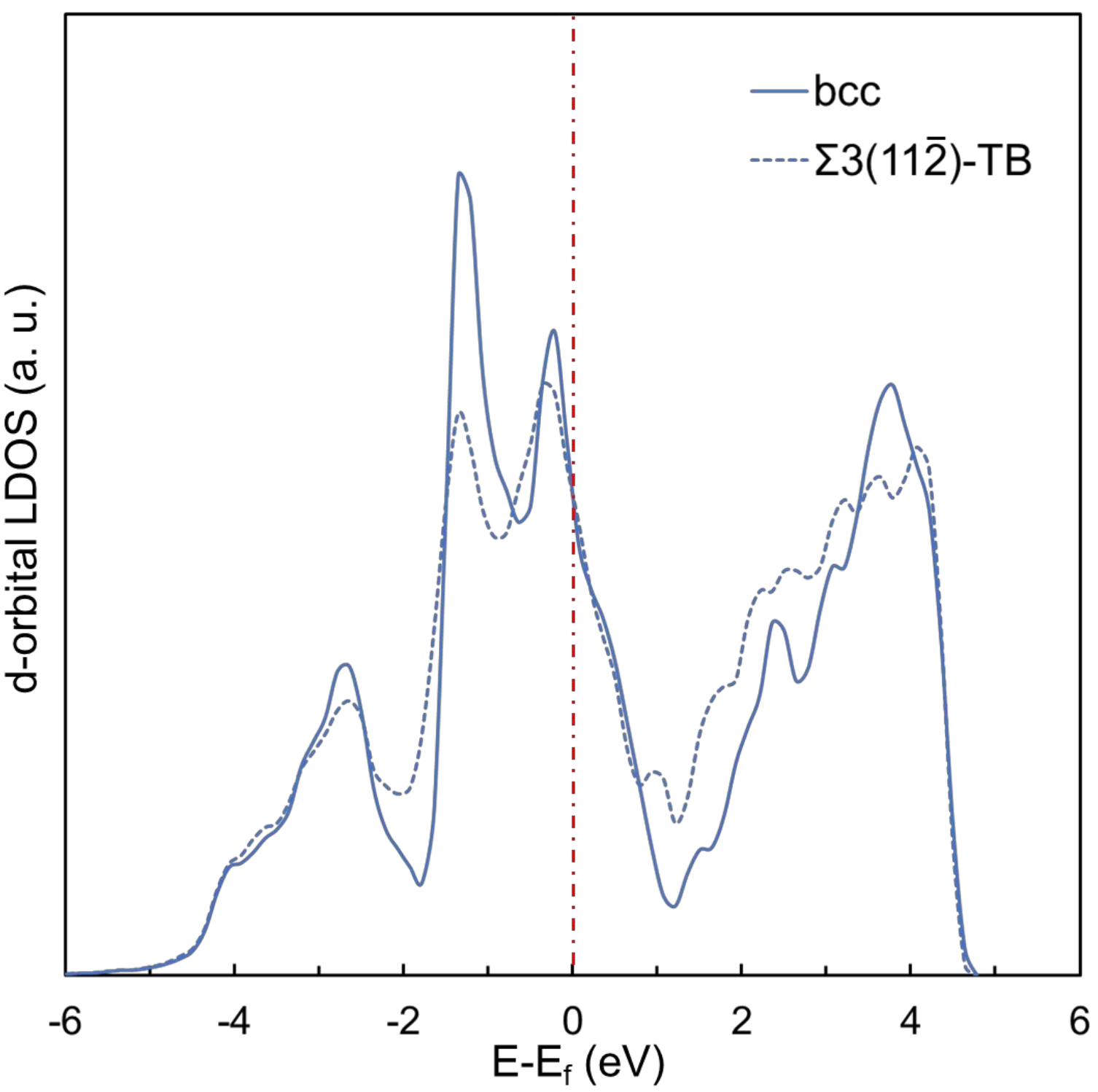

(b)

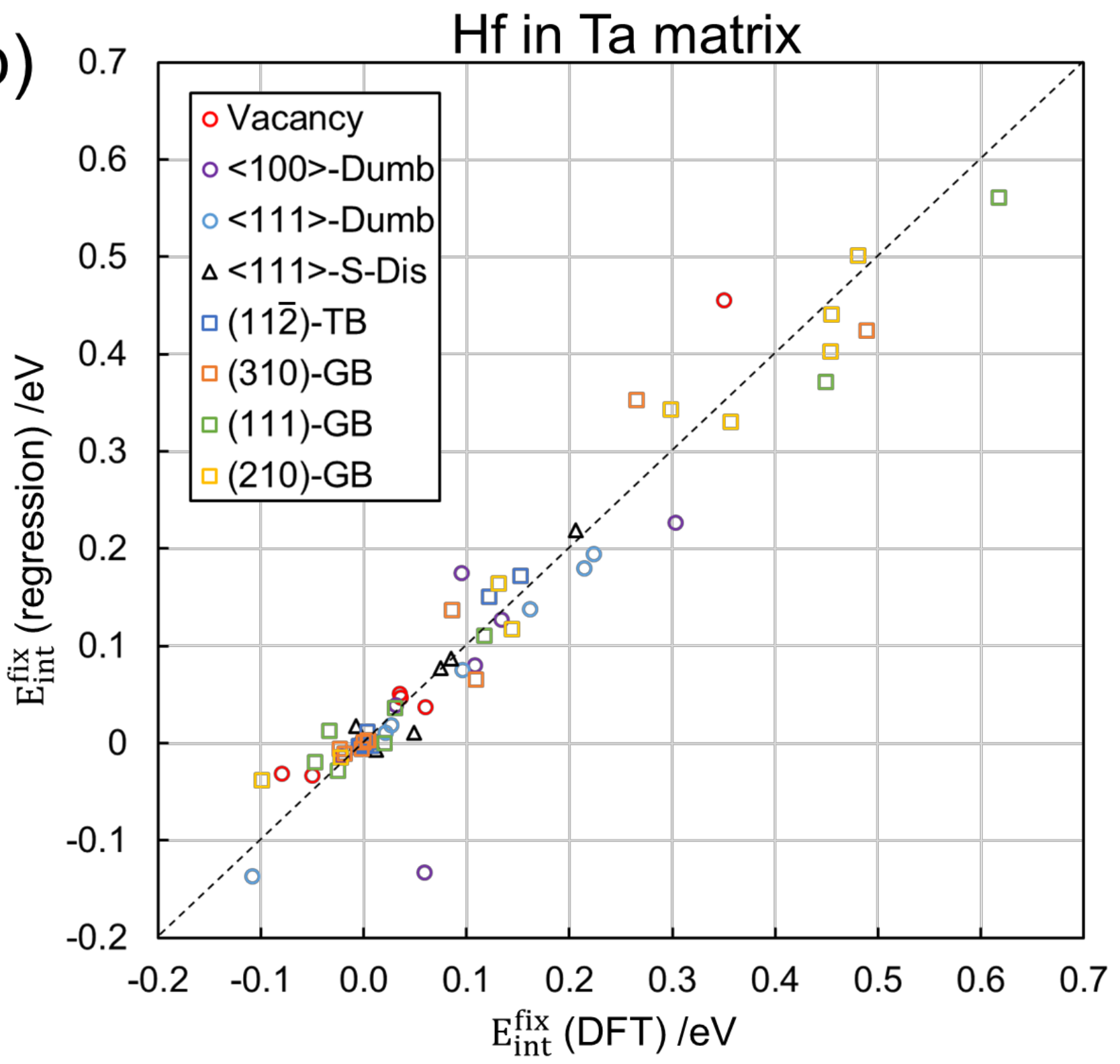
Hf in Ta matrix
Vacancy
<100>-Dumb
<111>-Dumb
<111>-S-Dis
(11$\bar{2}$)-TB
(310)-GB
(111)-GB
(210)-GB
$E_{int}^{fix}$ (regression) /eV
$E_{int}^{fix}$ (DFT) /eV
0.7
0.6
0.5
0.4
0.3
0.2
0.1
0
-0.1
-0.2
-0.2 -0.1 0.0 0.1 0.2 0.3 0.4 0.5 0.6 0.7

(c)

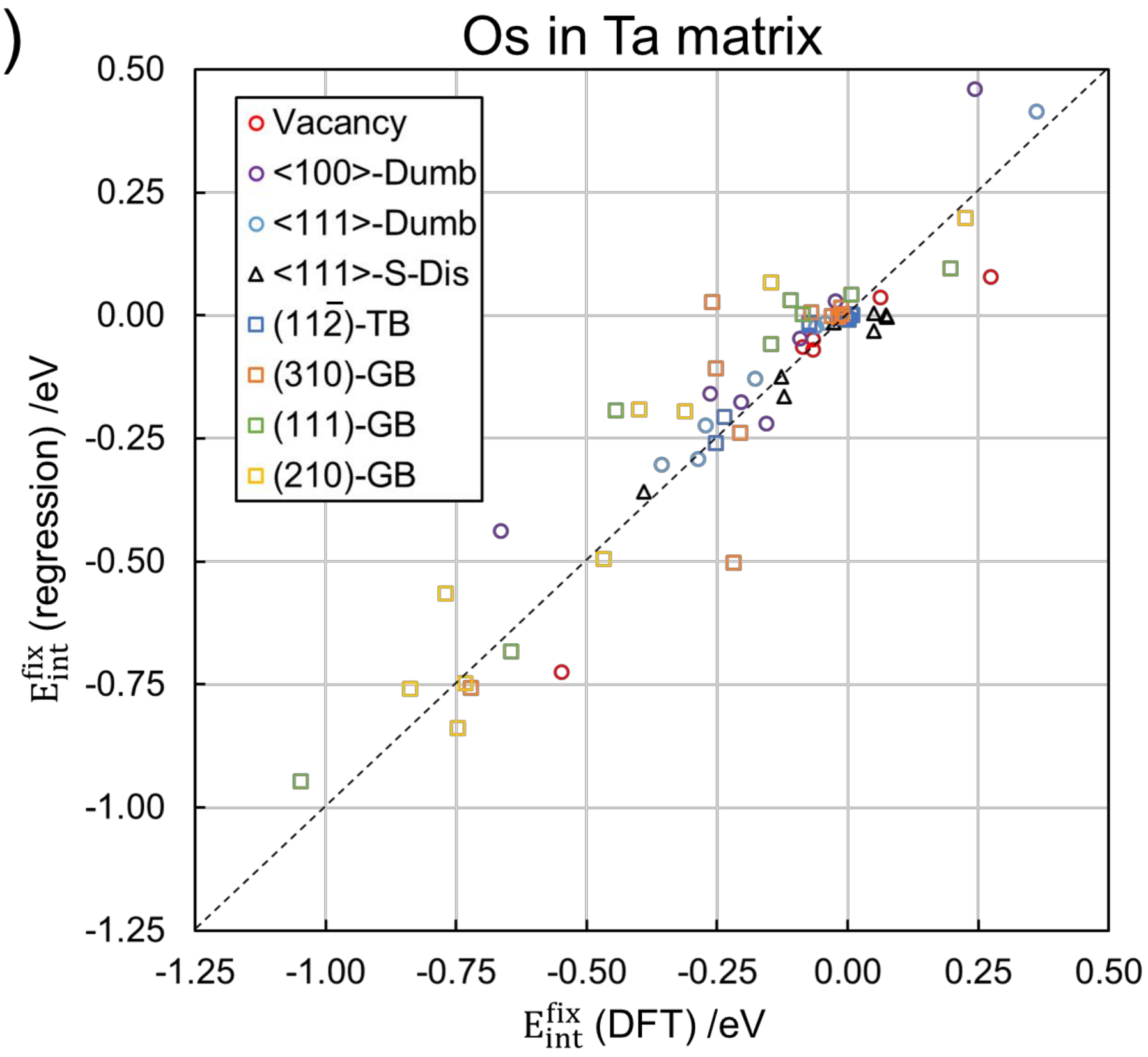
Os in Ta matrix
Vacancy
<100>-Dumb
<111>-Dumb
<111>-S-Dis
(11$\bar{2}$)-TB
(310)-GB
(111)-GB
(210)-GB
$E_{int}^{fix}$ (regression) /eV
$E_{int}^{fix}$ (DFT) /eV
0.50
0.25
0.00
-0.25
-0.50
-0.75
-1.00
-1.25
-1.25 -1.00 -0.75 -0.50 -0.25 0.00 0.25 0.50

**Figure 6. A schematic illustration of the electronic dependence of the regression coefficients.** The regression coefficients of the $\Delta dip$ and $x_{sp}$ term, $a_1$ and $a_2$, both have strong dependences on the $d$-orbital features of the solute elements. On the one hand, $a_1$ should have a positive sign when the position of Fermi Level ($E_f$) on the local $d$-band of solute atom is closer to the band edge compared to that of the matrix atom. In contrast, $a_1$ becomes negative when the position of $E_f$ on the local $d$-band of solute atom is closer to the minimum of the bcc pseudo-band gap. On the other hand, $a_2$ should have a positive sign when the solute atom has a larger d-orbital spatial extent than the matrix atom and vice versa. Therefore, based on the $d$-orbital features of the solute and matrix element, the solute-matrix pair studied in the present work can be classified into four different quadrants as shown in the figure.

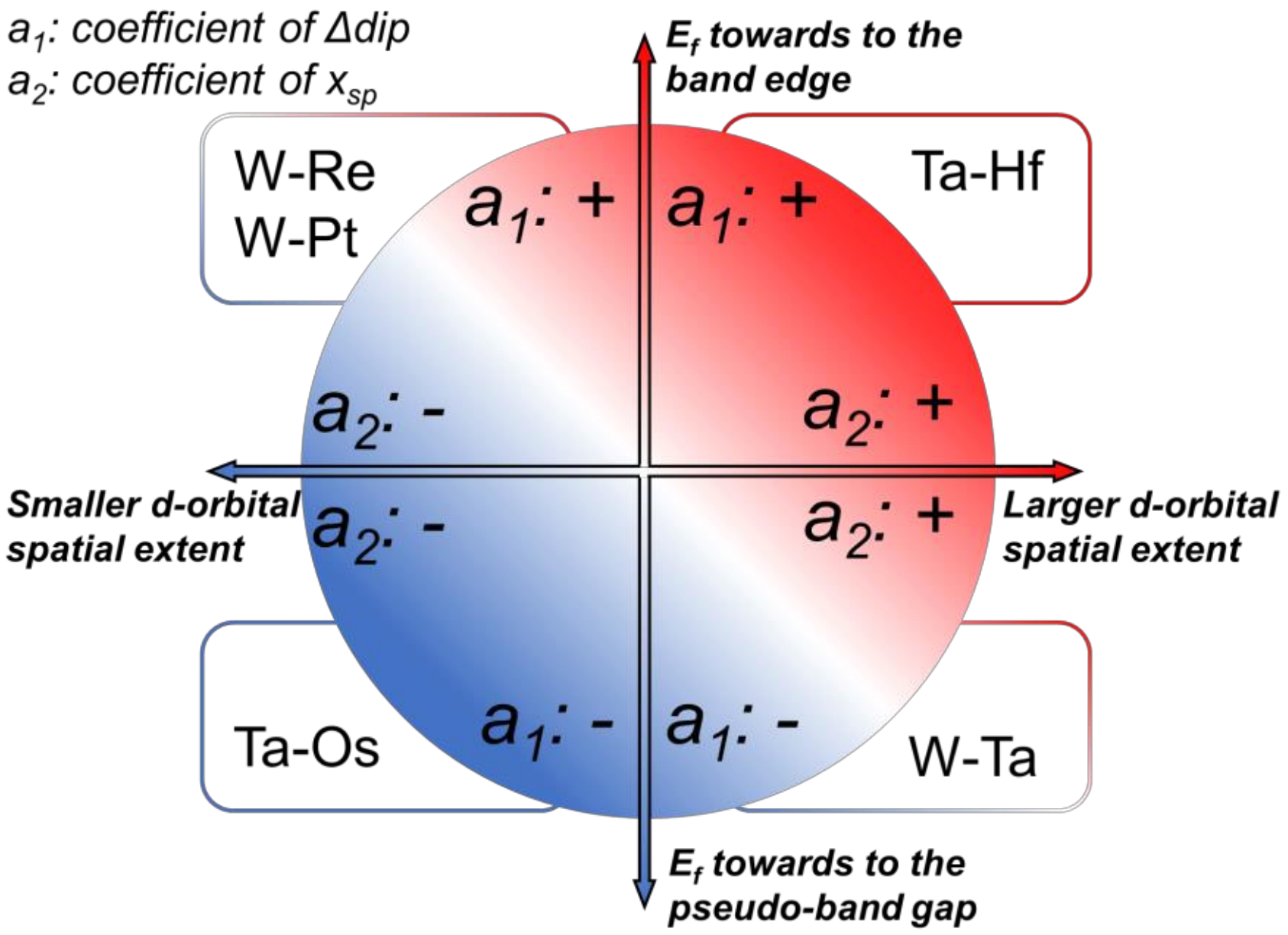

**Reference:**

1. Leyson, G. P. M., Curtin, W. A., Hector Jr, L. G. & Woodward, C. F. Quantitative prediction of solute strengthening in aluminium alloys. *Nature Materials* **9,** 750 (2010).

2. Wu, Z., Ahmad, R., Yin, B., Sandlöbes, S. & Curtin, W. A. Mechanistic origin and prediction of enhanced ductility in magnesium alloys. *Science* **359,** 447 LP-452 (2018).

3. Nie, J. F., Zhu, Y. M., Liu, J. Z. & Fang, X. Y. Periodic segregation of solute atoms in fully coherent twin boundaries. *Science* **340,** 957-960 (2013).

4. Trinkle, D. R. & Woodward, C. The chemistry of deformation: how solutes soften pure metals. *Science* **310,** 1665–1667 (2005).

5. Wakeda, M. *et al.* Chemical misfit origin of solute strengthening in iron alloys. *Acta Materialia* **131,** 445–456 (2017).

6. Xu, A. *et al.* Ion-irradiation-induced clustering in W-Re and W-Re-Os alloys: A comparative study using atom probe tomography and nanoindentation measurements. *Acta Materialia* **87,** 121–127 (2015).

7. Davidson, D. L. & Brotzen, F. R. Plastic deformation of molybdenum-rhenium alloy crystals. *Acta Metallurgica* **18,** 463–470 (1970).

8. Romaner, L., Ambrosch-Draxl, C. & Pippan, R. Effect of rhenium on the dislocation core structure in tungsten. *Physical review letters* **104,** 195503 (2010).

9. Hu, Y. J. *et al.* Solute-induced solid-solution softening and hardening in bcc tungsten. *Acta Materialia* **141,** 304-316 (2017).

10. Rodney, D., Ventelon, L., Clouet, E., Pizzagalli, L. & Willaime, F. Ab initio modeling of dislocation core properties in metals and semi-conductors. *Acta Materialia* **124**, 633-659 (2016).

11. Chookajorn, T., Murdoch, H. A. & Schuh, C. A. Design of stable nanocrystalline alloys. *Science* **337,** 951–954 (2012).

12. Argon, A. S. *Strengthening mechanisms in crystal plasticity*. (Oxford University Press Oxford, 2008).

13. Wolverton, C. Solute–vacancy binding in aluminum. *Acta Materialia* **55,** 5867–5872 (2007).

14. Clouet, E., Garruchet, S., Nguyen, H., Perez, M. & Becquart, C. S. Dislocation interaction with C in α-Fe: A comparison between atomic simulations and elasticity theory. *Acta Materialia* **56,** 3450–3460 (2008).

15. Shin, D. & Wolverton, C. First-principles study of solute–vacancy binding in magnesium. *Acta Materialia* **58,** 531–540 (2010).

16. Shang, S. L., Hargather, C. Z., Fang, H. Z., Wang, Y., Yong, D. & Liu, Z. K. Effects of alloying element and temperature on the stacking fault energies of dilute Ni-base superalloys. *Journal of Physics: Condensed Matter* **24,** 505403

(2012).

17. Naghavi, S. S., Hegde, V. I., Saboo, A. & Wolverton, C. Energetics of cobalt alloys and compounds and solute–vacancy binding in fcc cobalt: A first-principles database. *Acta Materialia* **124,** 1–8 (2017).

18. Ohnuma, T., Soneda, N. & Iwasawa, M. First-principles calculations of vacancy-solute element interactions in body-centered cubic iron. *Acta Materialia* **57,** 5947–5955 (2009).

19. Kong, X. S. *et al.* First-principles calculations of transition metal–solute interactions with point defects in tungsten. *Acta Materialia* **66,** 172–183 (2014).

20. Medvedeva, N. I., Gornostyrev, Y. N. & Freeman, A. J. Electronic origin of solid solution softening in bcc molybdenum alloys. *Physical review letters* **94,** 136402 (2005).

21. Wu, X. *et al.* First-principles determination of grain boundary strengthening in tungsten: Dependence on grain boundary structure and metallic radius of solute. *Acta Materialia* **120,** 315–326 (2016).

22. Shi, S., Zhu, L., Zhang, H., Sun, Z. & Ahuja, R. Mapping the relationship among composition, stacking fault energy and ductility in Nb alloys: A first-principles study. *Acta Materialia* **144,** 853–861 (2018).

23. Zhang, X. *et al.* Effects of solute size on solid-solution hardening in vanadium alloys: A first-principles calculation. *Scripta Materialia* **100,** 106–109 (2015).

24. Pettifor, D. G. *Bonding and structure of molecules and solids*. (Oxford University Press, 1995).

25. Sutton, A. P. *Electronic structure of materials*. (Clarendon Press, 1993).

26. Dezerald, L. *et al.* Ab initio modeling of the two-dimensional energy landscape of screw dislocations in bcc transition metals. *Physical Review B* **89,** 024104 (2014).

27. Drautz, R. & Pettifor, D. G. Valence-dependent analytic bond-order potential for transition metals. *Physical Review B* **74,** 174117 (2006).

28. Andersen, O. K. Linear methods in band theory. *Physical Review B* **12,** 3060 (1975).

29. De Jong, M. *et al.* Electronic origins of anomalous twin boundary energies in hexagonal close packed transition metals. *Physical Review Letters* **115,** 065501 (2015).

30. Freeman, J. B. & Dale, R. Assessing bimodality to detect the presence of a dual cognitive process. *Behavior Research Methods* **45,** 83–97 (2013).

31. Hartigan, J. A. & Hartigan, P. M. The Dip Test of Unimodality. *The Annals of Statistics* **13,** 70–84 (1985).

32. Hodges, L., Ehrenreich, H. & Lang, N. D. Interpolation scheme for band structure of noble and transition metals: ferromagnetism and neutron diffraction in Ni. *Physical Review* **152,** 505 (1966).

33. Pettifor, D. G. Theory of energy bands and related properties of 4d transition metals. III. s and d contributions to the equation of state. *Journal of Physics F: Metal Physics* **8,** 219 (1978).

34. Pettifor, D. G. Theory of energy bands and related properties of 4d transition metals. I. Band parameters and their volume dependence. *Journal of Physics F: Metal Physics* **7,** 613 (1977).

35. Mueller, F. M. Combined interpolation scheme for transition and noble metals. *Physical Review* **153**, 659 (1967).

36. Pettifor, D. G. Accurate resonance-parameter approach to transition-Metal band structure. *Physical Review B* **2,** 3031 (1970).

37. Lambert, R. M. & Pacchioni, G. *Chemisorption and Reactivity on Supported Clusters and Thin Films: Towards an Understanding of Microscopic Processes in Catalysis*. **331,** (Springer Science & Business Media, 2013).

38. Xin, H., Holewinski, A., Schweitzer, N., Nikolla, E. & Linic, S. Electronic structure engineering in heterogeneous catalysis: Identifying novel alloy catalysts based on rapid screening for materials with desired electronic properties. *Topics in Catalysis* **55,** 376–390 (2012).

39. Harrison, W. A. *Electronic structure and the properties of solids: the physics of the chemical bond*. (Courier Corporation, 2012).

40. Steinhardt, P. J., Nelson, D. R. & Ronchetti, M. Bond-orientational order in liquids and glasses, *Physical Review B* **28**, 784 (1983).

41. Gomberg, J. A., Medford, A. J. & Kalidindi, S. R. Extracting knowledge from molecular mechanics simulations of grain boundaries using machine learning. *Acta Materialia* **133,** 100–108 (2017).

42. Mueller, T., Kusne, A. G. & Ramprasad, R. Machine learning in materials science: Recent progress and emerging applications. *Reviews in Computational Chemistry* **29,** 186–273 (2016).

43. Dezerald, L., Proville, L., Ventelon, L., Willaime, F. & Rodney, D. First-principles prediction of kink-pair activation enthalpy on screw dislocations in bcc transition metals: V, Nb, Ta, Mo, W, and Fe. *Physical Review B* **91,** 94105 (2015).

44. Han, J., Thomas, S. L. & Srolovitz, D. J. Grain-Boundary Kinetics: A Unified Approach. *Progress in Materials Science* **98**, 386-476 (2018).

45. Hammer, B., Morikawa, Y. & Nørskov, J. K. CO chemisorption at metal surfaces and overlayers. *Physical review letters* **76,** 2141 (1996).

46. Hammer, B. & Nørskov, J. K. in *Advances in catalysis* **45,** 71–129 (Elsevier, 2000).

47. Hume-Rothery, W., & Raynor, G. V. The structure of metals and alloys. *The Institute of Metals,* London (1962).

48. Tanaka, I., Rajan, K. & Wolverton, C. Data-centric science for materials innovation. *MRS Bulletin* **43,** 659–663 (2018).

49. Blöchl, P. E. Projector augmented-wave method. *Physical Review B* **50,** 17953 (1994).

50. Perdew, J. P., Burke, K. & Ernzerhof, M. Generalized gradient approximation made simple. *Physical Review Letters* **77,** 3865–3868 (1996).

51. Kresse, G. & Furthmüller, J. Efficient iterative schemes for ab initio total-energy calculations using a plane-wave basis set. *Physical Review B* **54,** 011169 (1996).

52. Methfessel, M. & Paxton, A. T. High-precision sampling for Brillouin-zone integration in metals. *Physical Review B* **40,** 3616–3621 (1989).

53. Yasi, J. A. & Trinkle, D. R. Direct calculation of the lattice Green function with arbitrary interactions for general crystals. *Physical Review E* **85,** 66706 (2012).

54. Trinkle, D. R. Lattice Green function for extended defect calculations: Computation and error estimation with long-range forces. *Physical Review B* **78,** 014110 (2008).

55. Mechler, F. A direct translation into MATLAB from the original FORTRAN code of Hartigan's Subroutine DIPTST algorithm. Retrieved from www.nicprice.net/diptest (2002)

# Supplementary materials

## Universal correlation between electronic factors and solute-defect interactions in bcc refractory metals

Yong-Jie Hu[1], Ge Zhao[2], Baiyu Zhang[3], Chaoming Yang[1], Zi-Kui Liu[4], Xiaofeng Qian[3], and Liang Qi[1]

[1]Department of Materials Science and Engineering, University of Michigan, Ann Arbor, Michigan 48109, USA

[2]Department of Statistics, Pennsylvania State University, State College, Pennsylvania, 16802, USA

[3]Department of Materials Science and Engineering, Texas A&M University, College Station, Texas 77843, USA

[4]Department of Materials Science and Engineering, Pennsylvania State University, State College, Pennsylvania 16802, USA

## 1. Electronic configurations of the pseudopotentials for DFT calculations

**Table S1.** The electronic configuration of the pseudopotential for each element used in the first-principles calculations. The electrons in the bracket are treated as inner-core electrons.

| | | | |
|---|---|---|---|
| Elements | Hf_pv | Ta_pv | W_pv |
| V_RHFIN | $([Xe]4f^{14})5p^{6}5d^{3}6s^{1}$ | $([Xe]4f^{14})5p^{6}5d^{4}6s^{1}$ | $([Xe]4f^{14})5p^{6}5d^{5}6s^{1}$ |
| Elements | Re | Os | Ir |
| V_RHFIN | $([Xe]4f^{14})5d^{6}6s^{1}$ | $([Xe]4f^{14})5d^{7}6s^{1}$ | $([Xe]4f^{14})5d^{8}6s^{1}$ |
| Elements | Pt | | |
| V_RHFIN | $([Xe]4f^{14})5d^{9}6s^{1}$ | | |

## 2. Atomistic geometries of the supercells used for first-principles calculations

In the following paragraphs, the input size and geometry of the simulation supercells used for the first-principles calculations are described in detail along with its k-point mesh settings.

**Vacancy:** A bcc 4×4×4 supercell is used to model the electronic structures of the mono-vacancy defect and the corresponding solute-vacancy interactions in bcc Ta and W. One vacancy per supercell is introduced by removing a single Ta or W atom, as shown in Fig. S1a. The atomic sites chosen for the solute occupation are labeled by numbers in Fig. S1a. A larger number, n, means a longer pairing distance between the solute atom and defect center. The atomic site labeled with "*ref*" serves as the reference site for calculating the solute-defect interaction energy. A 4×4×4 grid is used for the k-point mesh in the first-principles calculations.

**Self-interstitial atom defects (SIA):** In the present work, two types of high symmetry SIA defects are studied, namely the <100>- and <111>-dumbbell structures. The initial input supercell geometries of the two defects are shown in Figs. S1b and 1c, respectively. The solute-substitutional sites are also labeled following the same notation as for the vacancy-solute interactions. Based on the supercell size (a bcc 4×4×4 supercell), a 4×4×4 grid is used for the k-point mesh in the first-principles calculations. Moreover, after relaxation, it is found that the <111>-dumbbell structure is actually transformed to the <111>-crowdion structure, which is consistent with the previous calculation results[1].

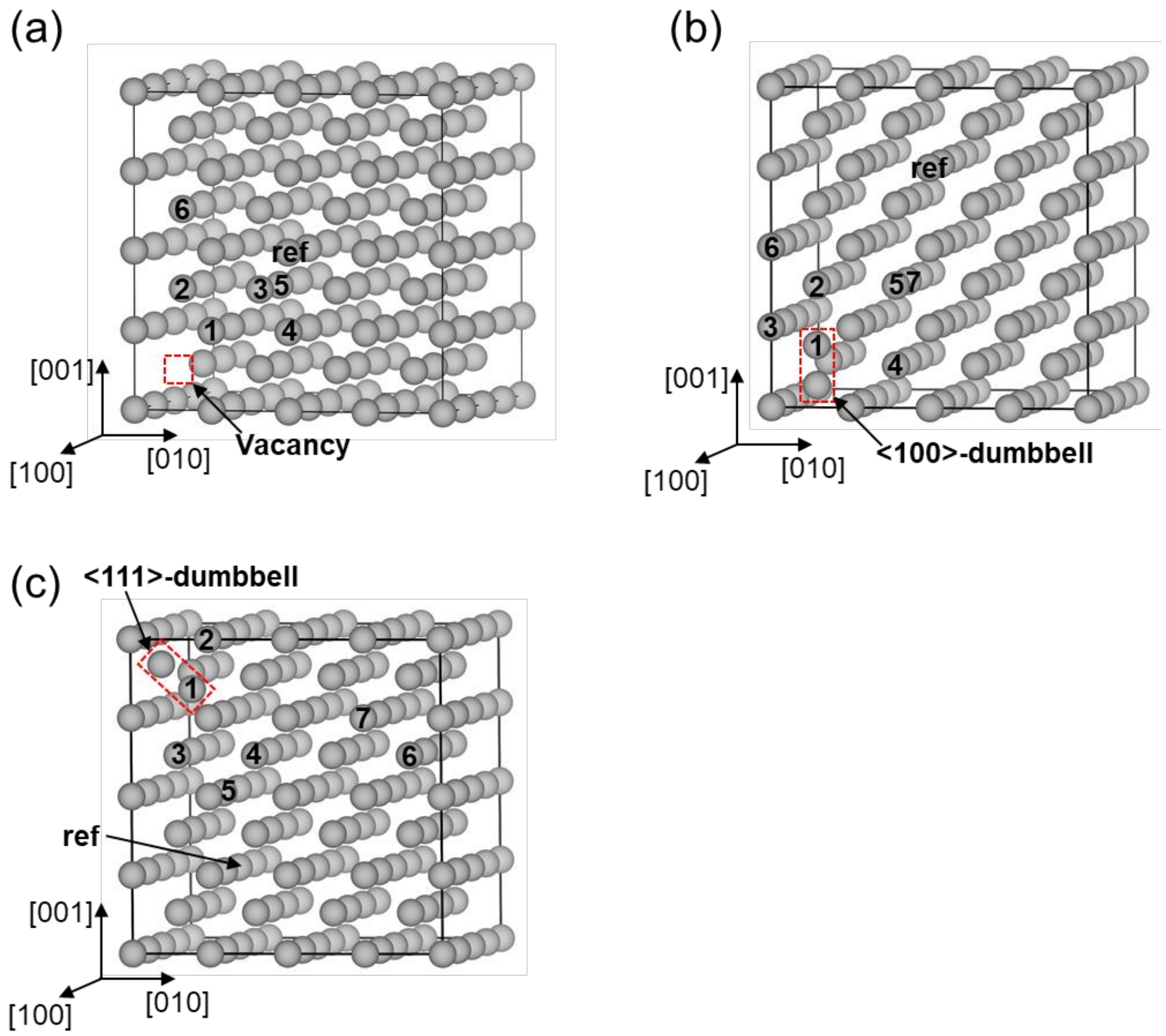

**Figure S1.** The input configurations of the supercells for the point defects. (a) Vacancy; (b) <100>-dumbbell; (c) <111>-dumbbell. The atomic sites for solute occupation are marked by numbers according to their relative distance to the defect center.

$\frac{1}{2}\langle\mathbf{111}\rangle$ **screw dislocation:** The supercell constructed for calculating the relaxed structure of the $\frac{1}{2}\langle 111\rangle$ screw dislocation has a geometry related to the bcc lattice index in terms of $5[11\bar{2}] \times 9[\bar{1}10] \times \frac{1}{2}[111]$. A single dislocation is placed in the center of the supercell, marked as a red dot in Fig. S2. The initial geometry of the dislocation is obtained by displacing the atoms according to the integral formulation of the anisotropic elastic displacement field[2]. The supercell contains 270 atoms and has a repeat length of one Burgers vector along the dislocation line direction. The supercell is relaxed via first-principles calculations using the flexible boundary condition method[3,4] with a k-point mesh of 1×1×16. The relaxation scheme is detailed described in our previous publication[5]. The relaxed supercell is then doubled along the z-axis (dislocation line direction) to make a new supercell. The latter is used for the calculations of solute-dislocation interactions. The purpose of this is to minimize the potential solute-solute interactions from the image interactions due to periodic boundary conditions in the DFT calculations.

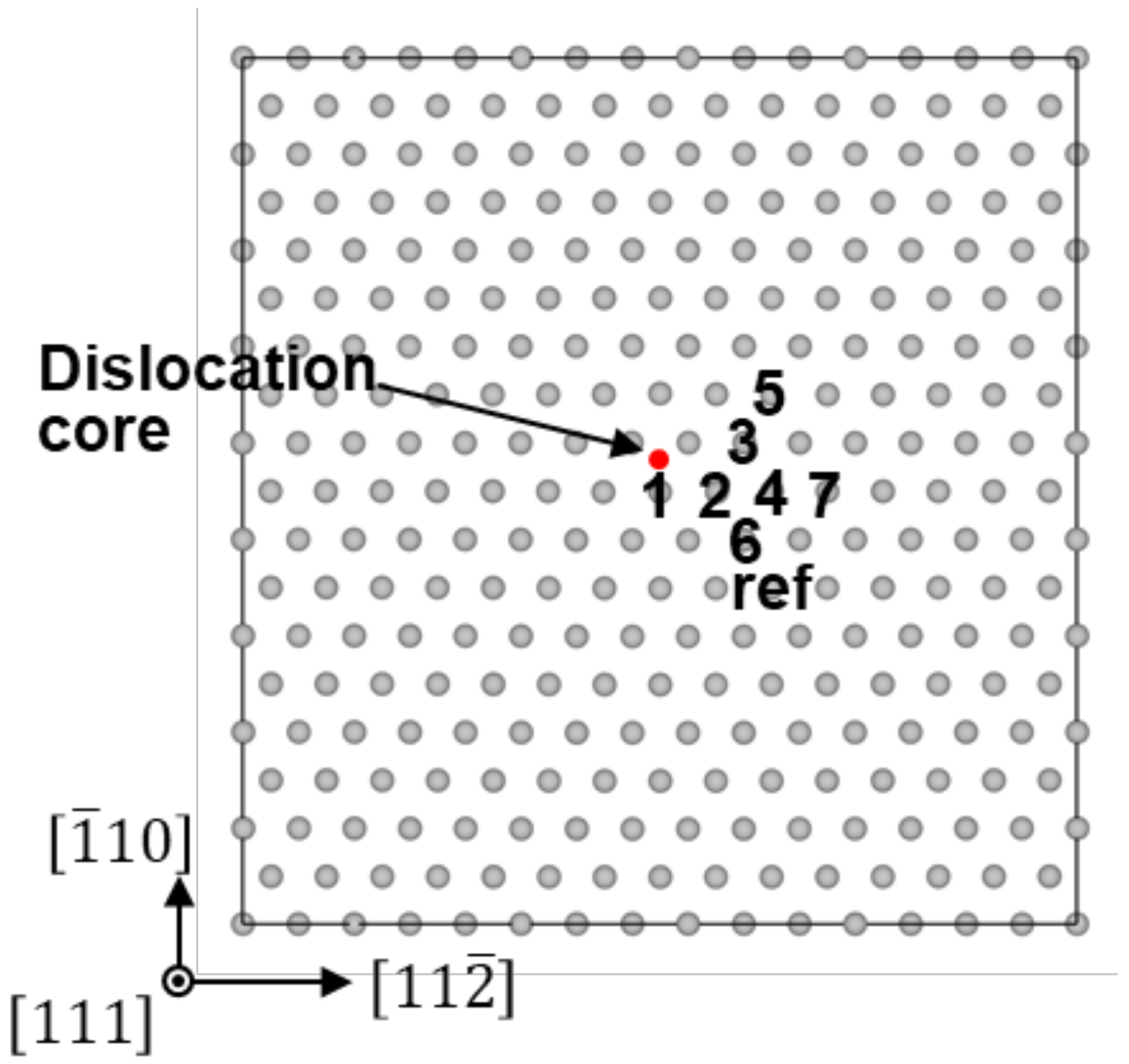

**Figure S2.** The input configurations of the supercell used for the calculations of the $\frac{1}{2}[111]$ screw dislocation. The atomic sites for solute occupation are marked by numbers according to their relative distance to the dislocation core.

**Twin/grain boundaries**: In the present work, first-principle calculations are performed for one twin boundary (TB) and three types of symmetric grain boundaries (GB) (i.e. $\Sigma3$ $(11\bar{2})$ TB, $\Sigma3$ (111) GB, $\Sigma5$ (310) GB and $\Sigma5$ (210) GB). The atomic configurations of these four interfacial defects are constructed using the coincidence site lattice (CSL) model. The input geometries of the supercells are shown in Figs. S3a, 3b, 3c and 3d for the $\Sigma3$ $(11\bar{2})$ TB, $\Sigma3$ (111) GB, $\Sigma5$ (310) GB and $\Sigma5$ (210) GB, respectively. As shown in Fig. S3, two interfaces are included in one supercell due to the periodic boundary conditions and are located apart from each other by around 20 Å to avoid the image interactions. The sizes of the supercells in the bcc lattice index are $6[11\bar{2}] \times [111] \times [1\bar{1}0]$, $7[111] \times [1\bar{1}0] \times [11\bar{2}]$, $4[310] \times [\bar{1}30] \times 2[001]$ and $6[210] \times [\bar{1}20] \times$ [001] for the $\Sigma3$ $(11\bar{2})$ TB, $\Sigma3$ (111) GB, $\Sigma5$ (310) GB and $\Sigma5$ (210) GB, respectively. Based on the supercell size, the grid of the k-point mesh for the first-principles calculations is set as 1×7×9, 1×9×5, 1×4×5 and 1×6×13. Similar to Figs. S1 and S2, the atomic sites chosen for the solute occupation are also labeled in Figs. S3. For each substitutional configuration, only one solute atom was introduced, giving an in-plane solute concentration of 50 at.%. Furthermore, it is noteworthy that the $\Sigma5$ (210) GB are found to loose its mirror symmetry after the structural relaxation. The similar phenomena were also reported in the previous works[6]. The relaxed structures of the $\Sigma5$ (210) GB in pure W and Ta configurations are shown in Figs. S3e and 3f, respectively.

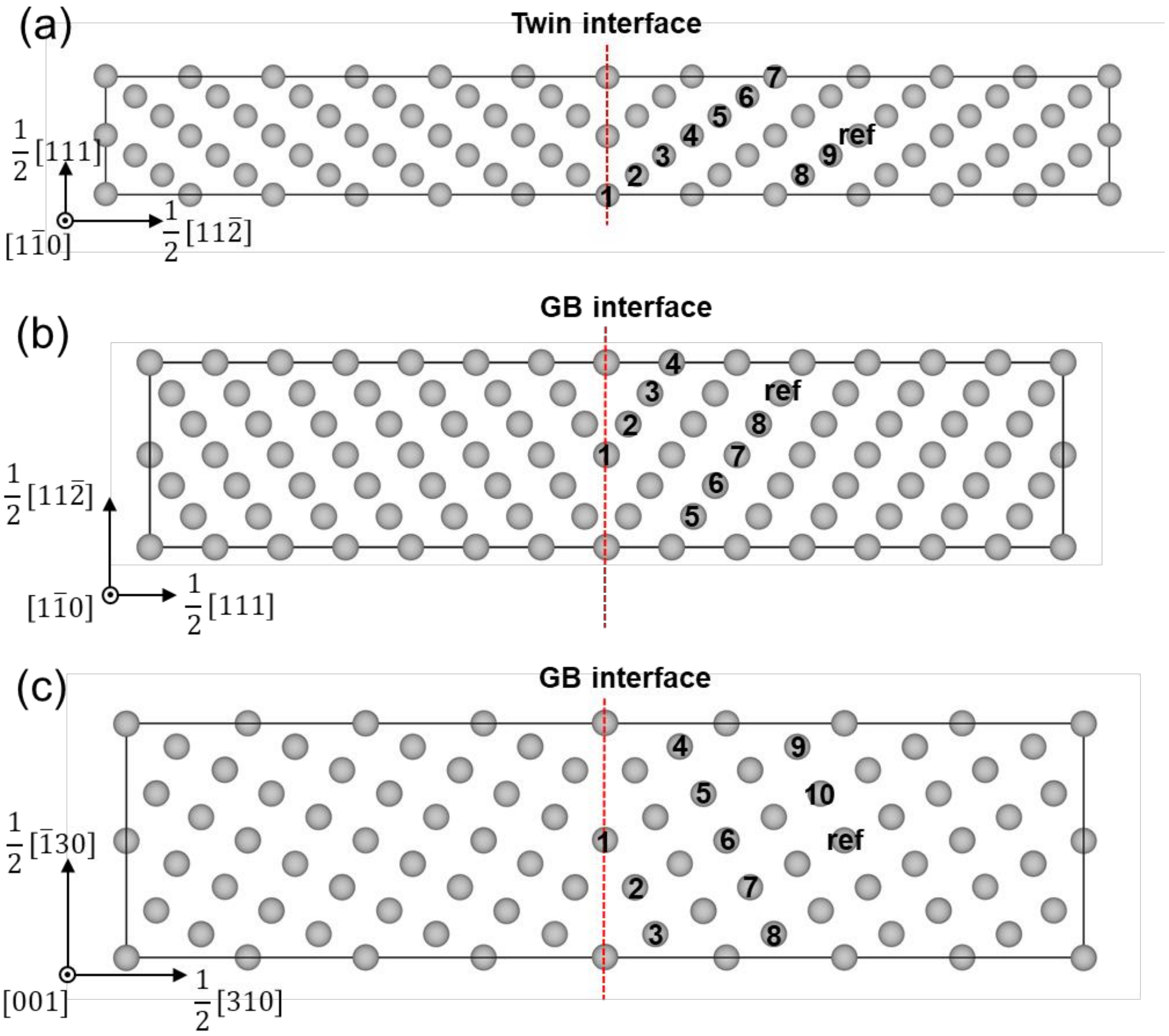

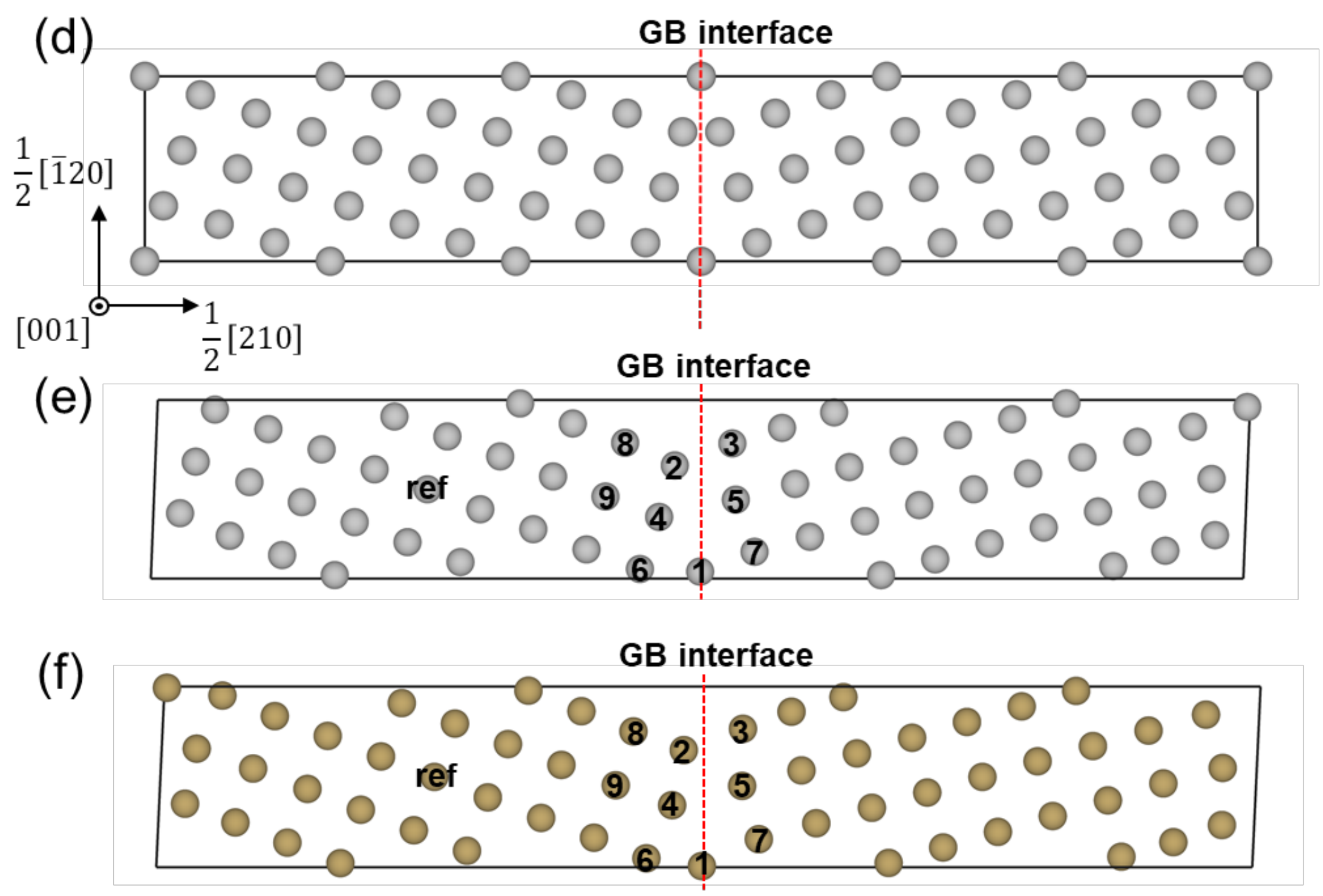

**Figure S3.** Configurations of the supercells for the TB and GBs. (a) Input geometry of the Σ3 $(11\bar{2})$ TB; (b) Input geometry of the Σ3 (110) GB; (c) Input geometry of the Σ5 (310) GB; (d) Input geometry of the Σ5 (210) GB; (e) Optimized geometry of the Σ5 (210) GB in pure W; (e) Optimized geometry of the Σ5 (210) GB in pure Ta. The atomic sites for solute occupation are marked by numbers according to their relative distance to the interface.

**Generalized stacking faults (GSF):** In the present work, the GSF of two typical slip systems in bcc crystal, namely <111>{110} and <111>{112}, are modeled via first-principles calculations. The supercell of the $[111](1\bar{1}0)$ GSF is constructed with periodicity in the slip plane with the x-axis along the [110] and the y-axis along [001] directions; along the z-axis there are 16 layers of the $(1\bar{1}0)$ planes with two atoms per plane, plus 12 Å of vacuum. Similarly, the supercell of the $[111](11\bar{2})$ GSF is composed of 12 layers of the $(11\bar{2})$ planes, plus 12 Å of vacuum. The x-axis and y-axis of the supercell are along the [111] and $[1\bar{1}0]$ directions, respectively. During GSF calculations, a slip vector $\vec{u}$ is applied by shifting the top half supercells rigidly along the [111] direction. Under the slip vector, first-principles calculations are performed by relaxing all atoms only along the z-axis. The atomic configurations of the GSF supercells before applying slip are shown in Figs. S4. Based on the supercell geometries, the k-mesh points are set as 9×13×1 and 9×15×1for the calculations of the $[111](1\bar{1}0)$

GSF and $[111](11\bar{2})$ GSF, respectively. The solute-defect interaction is only studied for the atomic site exactly on the fault plane. The solute-occupied defect and reference site are both marked out in Figs. S4. One solute is introduced to substitute one solvent atom in the relaxed GSF supercell, which gives an in-plane solute concentration of 50 at.%. In addition, the electronic structures of the GSF of the entire $(1\bar{1}0)$ plane, and its corresponding interactions with Re solute are modeled using the same supercell as it for the calculations of the $[111](1\bar{1}0)$ GSF. The $(1\bar{1}0)$ plane is screened by shifting the top half supercells rigidly along both the [001] and $[1\bar{1}0]$ directions.

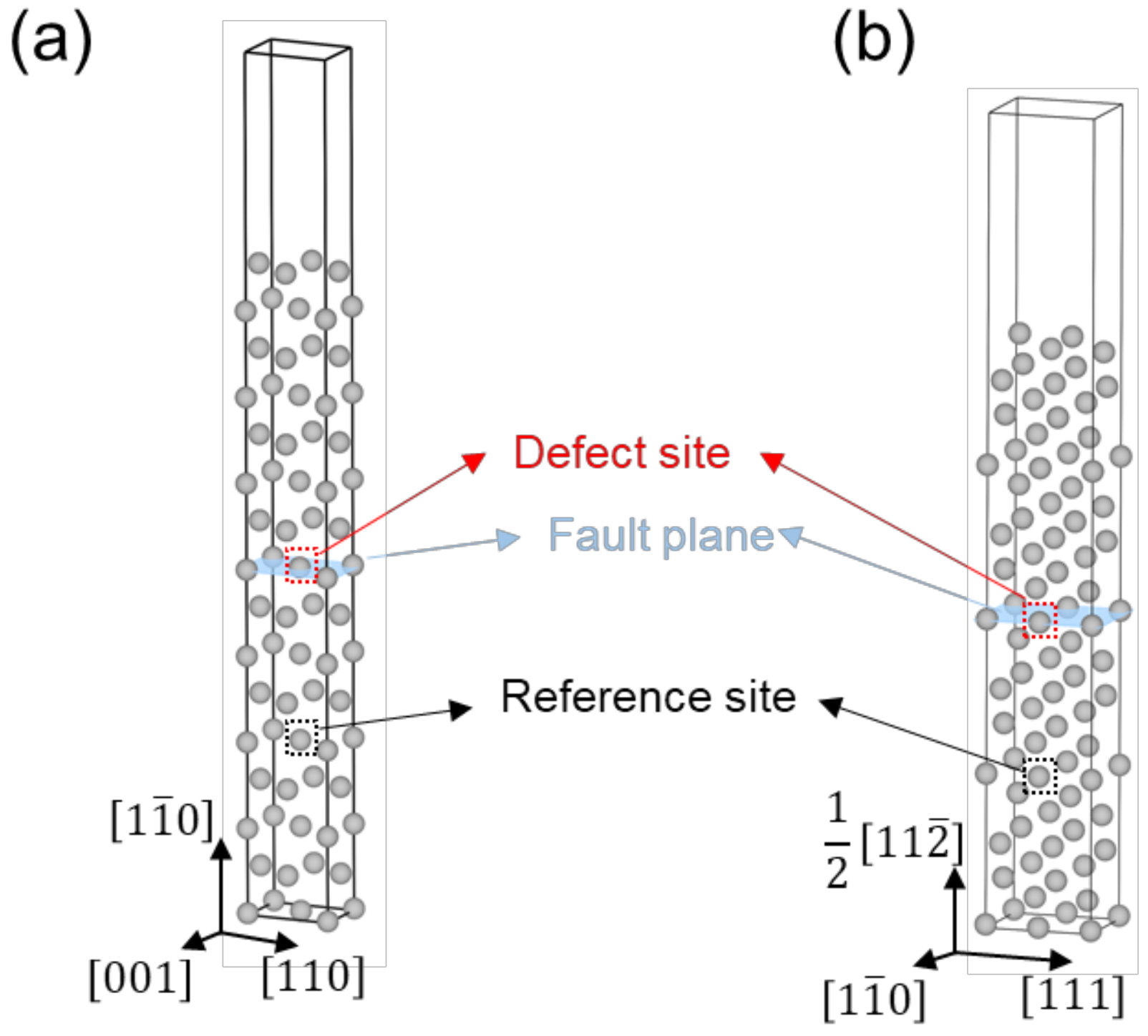

**Figure S4.** Configuration of the supercells used for the calculations of GSFs before applying slip. (a) $[111](1\bar{1}0)$ GSF; (b) $[111](11\bar{2})$ GSF. The atomic sites occupied by the solute atoms are marked out.

## 3. Results of $E_{int}^{fix}$, $\Delta dip$ and $x_{sp}$ from first-principles calculations

**Table S2.** Calculated $\Delta dip$ and $x_{sp}$ of the crystalline defects in bcc W along with their fixed-lattice interaction energies ($E_{int}^{fix}$) with the Ta, Re and Pt solutes. The results are listed for each defect site of interest, which is marked by numbers in Figs. S1-S3.

| Defect geometry | Site number | $\Delta dip$ | $x_{sp}$ | $E_{int}^{fix}$ (eV) | | |
|---|---|---|---|---|---|---|
| | | | | Ta | Re | Pt |
| Mono-vacancy | 1 | 0.01837 | 0.07173 | -0.0998 | 0.2263 | 0.8757 |
| | 2 | 0.00653 | -0.04608 | -0.1840 | 0.1806 | 0.2236 |
| | 3 | 0.00183 | -0.00674 | 0.0001 | 0.0149 | 0.0142 |
| | 4 | 0.00104 | -0.00289 | -0.0022 | 0.0212 | 0.0981 |
| | 5 | 0.00459 | -0.00420 | -0.0297 | 0.0310 | 0.0903 |
| | 6 | -0.00261 | 0.00248 | 0.0269 | -0.0286 | -0.0628 |
| <100>-dumbbell | 1 | 0.03006 | -0.34620 | -0.7279 | 0.9235 | 2.7096 |
| | 2 | 0.02376 | -0.04405 | -0.1812 | 0.3015 | 0.8689 |
| | 3 | 0.01319 | -0.00895 | -0.1023 | 0.2219 | 0.9095 |
| | 4 | -0.00174 | 0.09142 | 0.2147 | -0.1238 | 0.2453 |
| | 5 | 0.00067 | 0.02582 | 0.0592 | -0.0150 | 0.2429 |
| | 6 | 0.00575 | -0.00652 | -0.0664 | 0.0806 | 0.2829 |
| | 7 | 0.00157 | -0.01575 | -0.0525 | 0.0394 | 0.0743 |
| <111>-dumbbell | 1 | 0.03053 | -0.26858 | -0.6675 | 0.7907 | 2.1882 |
| | 2 | 0.01170 | 0.02660 | -0.0372 | 0.1561 | 0.7601 |
| | 3 | 0.00703 | 0.01912 | -0.0067 | 0.0959 | 0.5490 |
| | 4 | 0.00977 | 0.02092 | -0.0414 | 0.1164 | 0.5651 |
| | 5 | 0.00524 | 0.01162 | -0.0127 | 0.0731 | 0.3682 |
| | 6 | 0.00231 | 0.00240 | -0.0509 | 0.0505 | 0.1511 |
| | 7 | 0.00193 | 0.00528 | -0.0383 | 0.0469 | 0.1490 |
| $\frac{1}{2}\langle 111\rangle$ screw dislocation | 1 | 0.01414 | 0.00405 | -0.0445 | 0.2098 | 1.1231 |
| | 2 | 0.00703 | -0.00135 | -0.0209 | 0.0874 | 0.5958 |
| | 3 | 0.00654 | 0.00309 | 0.0020 | 0.0407 | 0.4744 |
| | 4 | 0.00156 | -0.00090 | -0.0068 | 0.0135 | 0.0447 |
| | 5 | 0.00329 | -0.00167 | -0.0250 | 0.0396 | 0.2104 |
| | 6 | 0.00099 | -0.00022 | -0.0079 | 0.0177 | 0.0770 |
| | 7 | 0.00012 | 0.00073 | 0.0039 | -0.0041 | -0.0034 |
| | 8 | -0.00005 | 0.00020 | 0.0014 | -0.0012 | -0.0045 |
| $\Sigma 3(11\bar{2})$ TB | 1 | 0.00957 | 0.00651 | -0.0265 | 0.1793 | 0.5477 |
| | 2 | 0.01288 | -0.02002 | -0.1218 | 0.2396 | 0.7446 |
| | 3 | 0.00306 | -0.01764 | -0.0823 | 0.1057 | 0.1944 |
| | 4 | 0.00056 | -0.01923 | -0.0734 | 0.0661 | 0.0152 |
| | 5 | -0.00069 | -0.01059 | -0.0402 | 0.0295 | -0.0596 |
| | 6 | -0.00016 | -0.00409 | -0.0138 | 0.0152 | -0.0018 |
| | 7 | -0.00013 | -0.00110 | 0.0031 | 0.0045 | 0.0010 |
| | 8 | -0.00030 | -0.00028 | 0.0063 | -0.0002 | 0.0122 |
| | 9 | -0.00003 | -0.00022 | 0.0010 | -0.0003 | -0.0090 |
| $\Sigma 3(111)$ GB | 1 | 0.02312 | 0.15556 | 0.2010 | 0.1923 | 1.3579 |
| | 2 | 0.02090 | -0.15116 | -0.5423 | 0.6293 | 1.4170 |
| | 3 | 0.01092 | 0.06353 | -0.0848 | 0.2361 | 0.9214 |
| | 4 | 0.00496 | -0.01023 | -0.0651 | 0.1181 | 0.1813 |
| | 5 | 0.00197 | -0.03506 | -0.1109 | 0.1169 | -0.0280 |
| | 6 | 0.00264 | -0.01301 | -0.0773 | 0.0703 | 0.0737 |
| | 7 | 0.00217 | -0.00358 | -0.0317 | 0.0338 | -0.0315 |
| | 8 | 0.00096 | -0.00191 | -0.0503 | 0.0533 | 0.0382 |
| $\Sigma 5(310)$ GB | 1 | 0.02254 | 0.11231 | 0.1472 | 0.1079 | 0.8986 |
| | 2 | 0.03006 | -0.01120 | -0.2484 | 0.4839 | 1.5242 |
| | 3 | 0.01160 | -0.01821 | -0.1457 | 0.2291 | 0.7396 |
| | 4 | 0.00575 | -0.02215 | -0.1108 | 0.1100 | 0.2283 |
| | 5 | 0.00202 | -0.01255 | -0.0587 | 0.0697 | 0.1439 |
| | 6 | -0.00331 | -0.00348 | -0.0430 | -0.0009 | -0.1392 |
| | 7 | 0.00006 | -0.00322 | -0.0340 | 0.0154 | -0.0327 |
| | 8 | 0.00097 | -0.00026 | -0.0201 | 0.0133 | 0.0496 |
| | 9 | -0.00054 | -0.00041 | -0.0141 | -0.0118 | -0.0115 |
| | 10 | -0.00041 | -0.00283 | -0.0175 | -0.0069 | -0.0202 |
| Σ5(210) GB | 1 | 0.02840 | 0.10837 | -0.0312 | 0.3382 | 1.7014 |
| | 2 | 0.03093 | 0.02800 | -0.2155 | 0.5387 | 1.8451 |
| | 3 | 0.03013 | -0.16447 | -0.5172 | 0.6986 | 1.7709 |
| | 4 | 0.01966 | -0.00256 | -0.2326 | 0.3729 | 1.2487 |
| | 5 | 0.01558 | 0.13434 | 0.1485 | 0.0573 | 1.0862 |
| | 6 | 0.01203 | 0.00592 | -0.0824 | 0.1928 | 0.7215 |
| | 7 | 0.01179 | -0.01932 | -0.1470 | 0.2319 | 0.7663 |
| | 8 | 0.00708 | -0.03905 | -0.1972 | 0.2176 | 0.5459 |
| | 9 | 0.00798 | -0.01194 | -0.0890 | 0.1353 | 0.3926 |

**Table S3.** Calculated $\Delta dip$ and $x_{sp}$ of the crystalline defects in bcc Ta along with their fixed-lattice interaction energies with the Hf and Os solutes. The results are listed for each defect site of interest, which is marked by numbers in Figs. S1-S3.

| **Defect geometry** | **Site number** | $\Delta dip$ | $x_{sp}$ | $E_{int}^{fix}$ **(eV)** | |
|---|---|---|---|---|---|
| | | | | Hf | Os |
| Mono-vacancy | 1 | 0.02509 | 0.13836 | 0.3505 | -0.5476 |
| | 2 | 0.00261 | -0.02318 | -0.0800 | 0.2739 |
| | 3 | 0.00403 | 0.00949 | 0.0362 | -0.0860 |
| | 4 | -0.00425 | -0.00215 | -0.0503 | 0.0627 |
| | 5 | 0.00450 | 0.01024 | 0.0348 | -0.0658 |
| | 6 | 0.00340 | 0.00683 | 0.0596 | -0.0663 |
| <100>-dumbbell | 1 | 0.03039 | -0.16101 | 0.0595 | 0.2425 |
| | 2 | 0.01849 | 0.02479 | 0.0952 | -0.1558 |
| | 3 | 0.01118 | 0.02542 | 0.1342 | -0.2050 |
| | 4 | 0.00073 | 0.10643 | 0.3034 | -0.6644 |
| | 5 | -0.00021 | 0.03927 | 0.1083 | -0.2639 |
| | 6 | 0.00450 | 0.00455 | 0.0311 | -0.0914 |
| | 7 | 0.00045 | -0.00784 | -0.0216 | -0.0230 |
| <111>-dumbbell | 1 | 0.02276 | -0.13813 | -0.1085 | 0.3606 |
| | 2 | 0.01141 | 0.05669 | 0.2239 | -0.3557 |
| | 3 | 0.00706 | 0.04369 | 0.1614 | -0.2729 |
| | 4 | 0.00908 | 0.05746 | 0.2143 | -0.2868 |
| | 5 | 0.00268 | 0.02742 | 0.0965 | -0.1774 |
| | 6 | 0.00227 | 0.00172 | 0.0271 | -0.0615 |
| | 7 | 0.00097 | 0.00226 | 0.0213 | -0.0426 |
| $\frac{1}{2}\langle 111\rangle$ | 1 | 0.01071 | 0.07100 | 0.2065 | -0.3910 |
| | 2 | 0.00082 | 0.03897 | 0.0853 | -0.1225 |
| screw dislocation | 3 | 0.00375 | 0.02493 | 0.0749 | -0.1278 |
| | 4 | 0.00034 | 0.00739 | -0.0074 | 0.0506 |
| | 5 | 0.00090 | 0.00229 | 0.0487 | -0.0273 |
| | 6 | -0.00128 | 0.00099 | 0.0115 | 0.0492 |
| | 7 | -0.00016 | 0.00091 | -0.0024 | 0.0753 |
| | 8 | 0.00001 | 0.00008 | -0.0064 | 0.0732 |
| $\Sigma3(11\bar{2})$ TB | 1 | 0.01164 | 0.04554 | 0.1527 | -0.2527 |
| | 2 | 0.01343 | 0.02931 | 0.1222 | -0.2363 |
| | 3 | -0.00301 | 0.00853 | 0.0016 | -0.0744 |
| | 4 | 0.00042 | 0.00415 | 0.0036 | -0.0740 |
| | 5 | -0.00137 | 0.00354 | 0.0011 | -0.0160 |
| | 6 | -0.00137 | 0.00326 | 0.0073 | -0.0043 |
| | 7 | -0.00090 | 0.00355 | 0.0076 | 0.0021 |
| | 8 | -0.00096 | 0.00132 | -0.0035 | 0.0093 |
| | 9 | -0.00054 | 0.00026 | -0.0044 | 0.0063 |
| $\Sigma3(111)$ GB | 1 | 0.02271 | 0.19686 | 0.6176 | -1.0476 |
| | 2 | 0.01844 | -0.05261 | -0.0330 | 0.1966 |
| | 3 | 0.00639 | 0.15796 | 0.4496 | -0.6445 |
| | 4 | 0.00328 | 0.04247 | 0.1174 | -0.4448 |
| | 5 | -0.00106 | -0.00605 | -0.0471 | -0.1087 |
| | 6 | 0.00189 | 0.01151 | 0.0300 | -0.1471 |
| | 7 | 0.00025 | -0.00079 | 0.0203 | -0.0861 |
| | 8 | -0.00195 | -0.00731 | -0.0251 | 0.0076 |
| $\Sigma5(310)$ GB | 1 | 0.01076 | 0.16921 | 0.4894 | -0.7209 |
| | 2 | 0.02844 | 0.07849 | 0.2651 | -0.2182 |
| | 3 | 0.00425 | 0.05210 | 0.0860 | -0.2069 |
| | 4 | 0.00225 | -0.01014 | -0.0234 | -0.2595 |
| | 5 | 0.00299 | 0.02196 | 0.1088 | -0.2534 |
| | 6 | -0.00085 | -0.00024 | -0.0018 | -0.0706 |
| | 7 | 0.00059 | -0.00153 | -0.0002 | -0.0117 |
| | 8 | -0.00068 | -0.00277 | -0.0188 | -0.0119 |
| | 9 | 0.00090 | -0.00125 | 0.0033 | -0.0298 |
| | 10 | 0.00044 | 0.00023 | 0.0044 | -0.0186 |
| $\Sigma5(210)$ GB | 1 | 0.02138 | 0.17245 | 0.4808 | -0.7458 |
| | 2 | 0.02658 | 0.07989 | 0.2989 | -0.4672 |
| | 3 | 0.02591 | -0.08966 | -0.0220 | 0.2253 |
| | 4 | 0.01723 | 0.15659 | 0.4545 | -0.7319 |
| | 5 | 0.00400 | 0.18039 | 0.4542 | -0.8383 |
| | 6 | 0.01920 | 0.01762 | 0.1310 | -0.3117 |
| | 7 | 0.01204 | 0.11985 | 0.3565 | -0.7699 |
| | 8 | -0.00089 | -0.01514 | -0.0986 | -0.1471 |
| | 9 | 0.00553 | 0.03848 | 0.1445 | -0.4000 |

## 4. Comparison between $E_{int}^{relax}$ and $E_{int}^{fix}$

To evaluate the effects of the lattice relaxation on the solute-defect interaction, the "*relaxed interaction energies*" ($E_{int}^{relax}$), which were obtained by full relaxation of atomic positions due to solute substitutions, are also calculated for a few of defect sites that have relatively strong interactions with the solutes. The difference between the relaxed ($E_{int}^{relax}$) and fixed-lattice interaction energies ($E_{int}^{fix}$) gives the energy change due to the relaxation of the defect lattice upon the solute substitution. As shown in Fig. 1, in the W alloys, the relative difference between $E_{int}^{relax}$ and $E_{int}^{fix}$ of the solute-dislocation interactions is small, which indicates that the interactions may mainly originate from the changes in the local electronic bonding environment near the defects compared to the bulk lattice. It is worth to further test whether the difference between $E_{int}^{relax}$ and $E_{int}^{fix}$ is also small or not for other types of solute-defect interactions in the Ta and W alloys. Here, the Ta-Hf and W-Pt systems are chosen as two samples, in which the relaxed interaction energy ($E_{int}^{relax}$) is calculated for a few defect sites that have relatively strong interactions with the solutes. The calculated results are summarized in Tables S4 and S5 for the Ta-Hf and W-Pt systems, respectively. The corresponding values of $E_{int}^{fix}$ are also included in the tables for comparison. As shown in Tables S4 and S5, the difference between $E_{int}^{relax}$ and $E_{int}^{fix}$ is generally small for various types of solute-defect interactions in the Ta-Hf and W-Pt alloys. Along with the results of Fig. 1, it suggests that the relaxation of the atomic positions generally has minor effects on the solute-defect interactions in the bcc refractory alloys. The interactions mainly originate from the variations in the local electronic structures due to the presence of the defect geometry.

**Table S4.** Calculated $E_{int}^{relax}$ and $E_{int}^{fix}$ of the interactions between Hf and the crystalline defects in bcc Ta. (unit: eV)

| | **Vacancy** | **<111>-dumbbell** | **Σ3 (11$\bar{2}$) twin boundary** | **Σ5 (310) grain boundary** |
|---|---|---|---|---|
| Site* | 1-nn | 2-nn | 1-nn | 1-nn |
| $E_{int}^{relax}$ | 0.388 | 0.213 | 0.138 | 0.499 |
| $E_{int}^{fix}$ | 0.351 | 0.223 | 0.152 | 0.489 |

*1-nn: the atomic site with first shortest distance to defect center; 2-nn: the atomic site with second shortest distance to defect center

**Table S5.** Calculated $E_{int}^{relax}$ and $E_{int}^{fix}$ of the interactions between Pt and the crystalline defects in bcc W. (unit: eV)

| | Vacancy | | $\Sigma3\ (11\bar{2})$ twin boundary | | Σ5 (310) grain boundary | |
|---|---|---|---|---|---|---|
| Site* | 1-nn | 2-nn | 1-nn | 2-nn | 1-nn | 2-nn |
| $E_{int}^{relax}$ | 0.947 | 0.184 | 0.754 | 0.921 | 0.851 | 1.675 |
| $E_{int}^{fix}$ | 0.876 | 0.224 | 0.548 | 0.745 | 0.899 | 1.524 |

*1-nn: the atomic site with first shortest distance to defect center; 2-nn: the atomic site with second shortest distance to defect center

## 5. Correlation between the bimodality of LDOS and the solute-GSF interactions in bcc W

The variations in the electronic structure of the atom on the fault plane of the GSFs in pure bcc W are also investigated via first-principles calculations. $\Delta dip$ and $x_{sp}$ of the W atom on the fault plane is calculated as a function of the relative displacement of the $[111](1\bar{1}0)$ and $[111](11\bar{2})$ GSF, and summarized in Table S6 below.

**Table S6.** Calculated $\Delta dip$ and $x_{sp}$ of the W atom on the fault plane of the $[111](1\bar{1}0)$ and $[111](11\bar{2})$ GSF with different slip displacements, along with the fixed-lattice interaction energies when the W atom is substituted by the Re or Pt solutes. The value of the relative displacement along slip direction is relative to the length of Burgers vector, which is 1/2[111].

| GSF geometry | Relative displacement | $\Delta dip$ | $x_{sp}$ | $E_{int}^{fix}$ (eV) | |
|---|---|---|---|---|---|
| | | | | Re | Pt |
| $[111](1\bar{1}0)$ | 0.1 | 0.00303 | 0.00916 | 0.0040 | 0.1590 |
| | 0.2 | 0.00874 | 0.01580 | 0.0609 | 0.4899 |
| | 0.3 | 0.01314 | 0.01694 | 0.1264 | 0.7502 |
| | 0.4 | 0.01504 | 0.01439 | 0.1827 | 0.9053 |
| | 0.5 | 0.01502 | 0.01545 | 0.2037 | 0.9556 |
| | 0.6 | 0.01504 | 0.01439 | 0.1827 | 0.9053 |
| | 0.7 | 0.01314 | 0.01694 | 0.1264 | 0.7502 |
| | 0.8 | 0.00874 | 0.01580 | 0.0609 | 0.4899 |
| | 0.9 | 0.00303 | 0.00916 | 0.0040 | 0.1590 |
| $[111](11\bar{2})$ | 0.1 | 0.00353 | 0.01020 | 0.0130 | 0.2223 |
| | 0.2 | 0.01050 | 0.01124 | 0.1195 | 0.8254 |
| | 0.3 | 0.01703 | -0.00545 | 0.2766 | 1.2738 |
| | 0.4 | 0.01727 | -0.00078 | 0.3158 | 1.3629 |
| | 0.5 | 0.01508 | 0.01327 | 0.2707 | 1.2009 |
| | 0.6 | 0.01495 | 0.01860 | 0.2220 | 1.0360 |
| | 0.7 | 0.01548 | 0.01934 | 0.2189 | 1.0368 |
| | 0.8 | 0.01339 | 0.01865 | 0.1723 | 0.9180 |

| | | | | |
|---|---|---|---|---|
| 0.9 | 0.00686 | 0.01554 | 0.0519 | 0.4331 |

As shown in Table S6, the creation of GSF results in variations in the local electronic structure of the atom on the fault plane, associated with an increase of the $\Delta dip$ parameter. In addition, $E_{int}^{fix}$ are also calculated for the Re and Pt solutes. Figs. S5a and 5b plotted the calculated $\Delta dip$ and $E_{int}^{fix}$ with respect to the relative displacement along the slip direction for the $[111](1\bar{1}0)$ and $[111](11\bar{2})$ GSF, respectively. As shown in Fig. S5a and 5b, the shape of the interaction-energy curves is very similar to that of the $\Delta dip$ curves, for both Re and Pt, indicating the strong correlation between the LDOS bimodality and the solute-defect interactions. Furthermore, it is found that this correlation can be extended to the defect geometries that artificially frozen in unstable states. Fig. S6a shows a contour plot of the $\Delta dip$ for the GSF of the entire $(1\bar{1}0)$ plane in pure W. Correspondingly, as shown in Fig. S6b, the contour plot of $E_{int}^{fix}$ of Re exhibits a similar gradient change as that of $\Delta dip$.

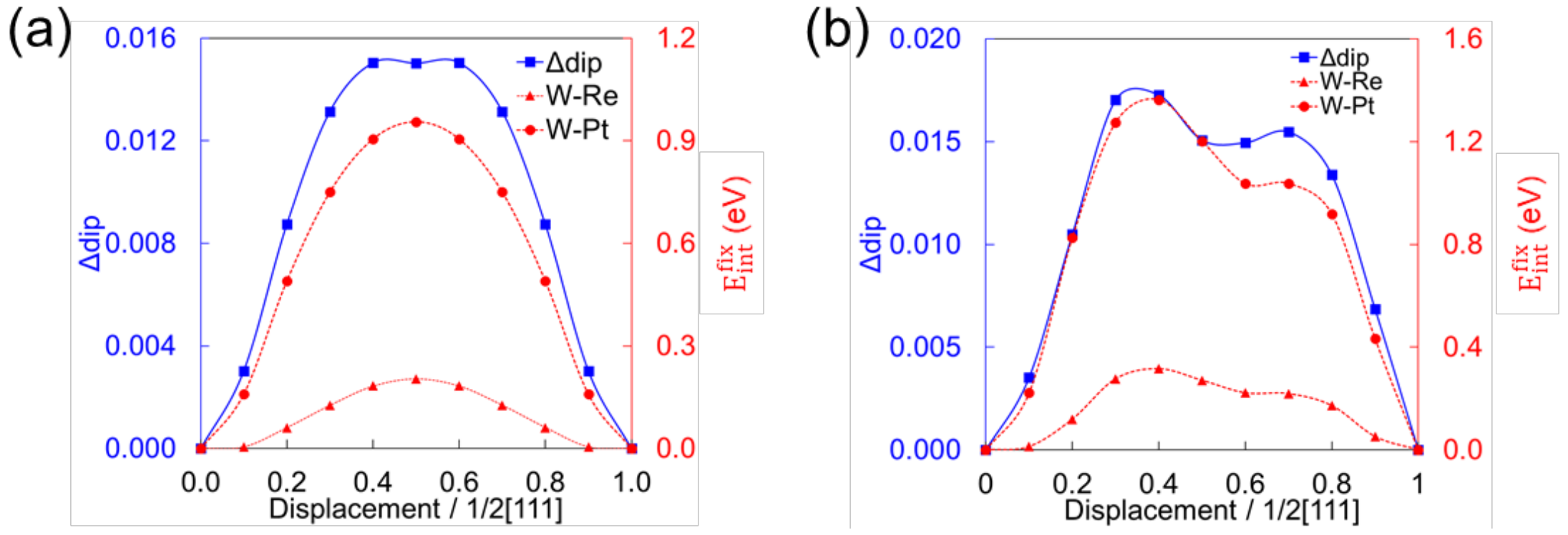

**Figure S5.** Changes in bimodalities of LDOSs ($\Delta dip$) of the atom on fault plane with respect to the relative displacement of the GSFs in pure W, along with the corresponding solute-defect interaction energies ($E_{int}^{fix}$) when the W atom is substituted by a solute atom (Re and Pt). (a) $[111](1\bar{1}0)$ GSF; (b) $[111](11\bar{2})$ GSF.

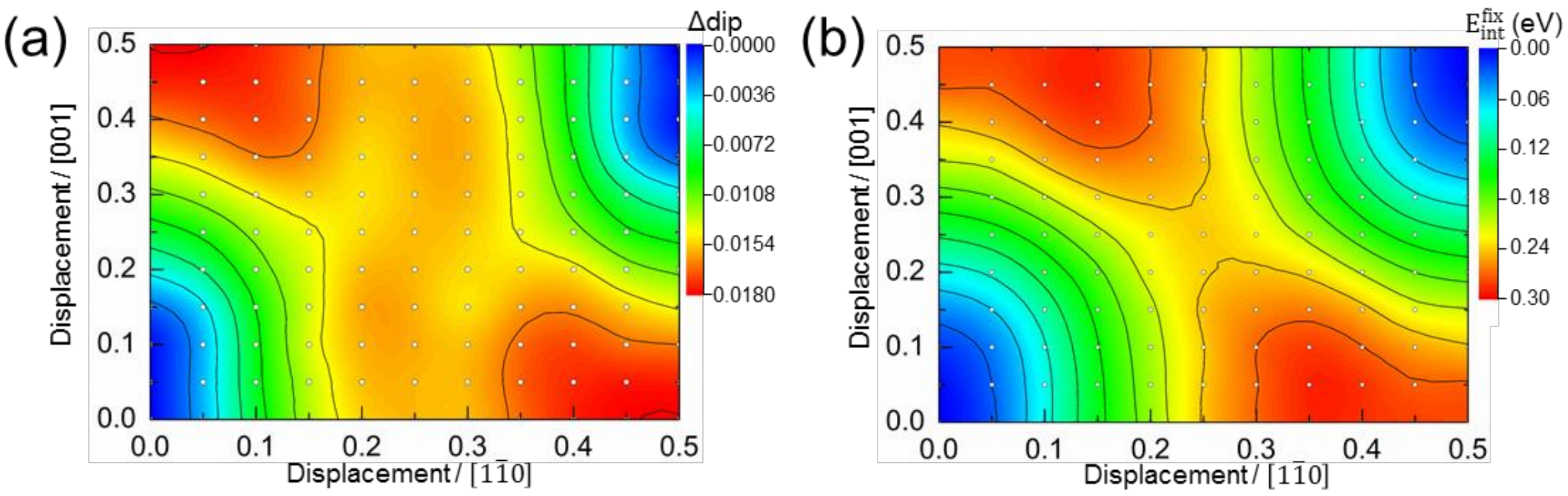

**Figure S6.** (a) Contour plot of $\Delta dip$ of the atom on the defect plane of the GSF of the entire $(1\bar{1}0)$ plane in pure W. (b) Contour plot of $E_{int}^{fix}$ when one W atom on the defect plane is substituted by a Re atom. The white dots in both figures represent the slip displacements where the first-principles calculations are performed. The coloring between each white dot is a linear-interpolation of the DFT results.

## 6. General correlation between $\Delta dip$ and $E_{int}^{fix}$ in the W-Re and W-Pt binary alloys

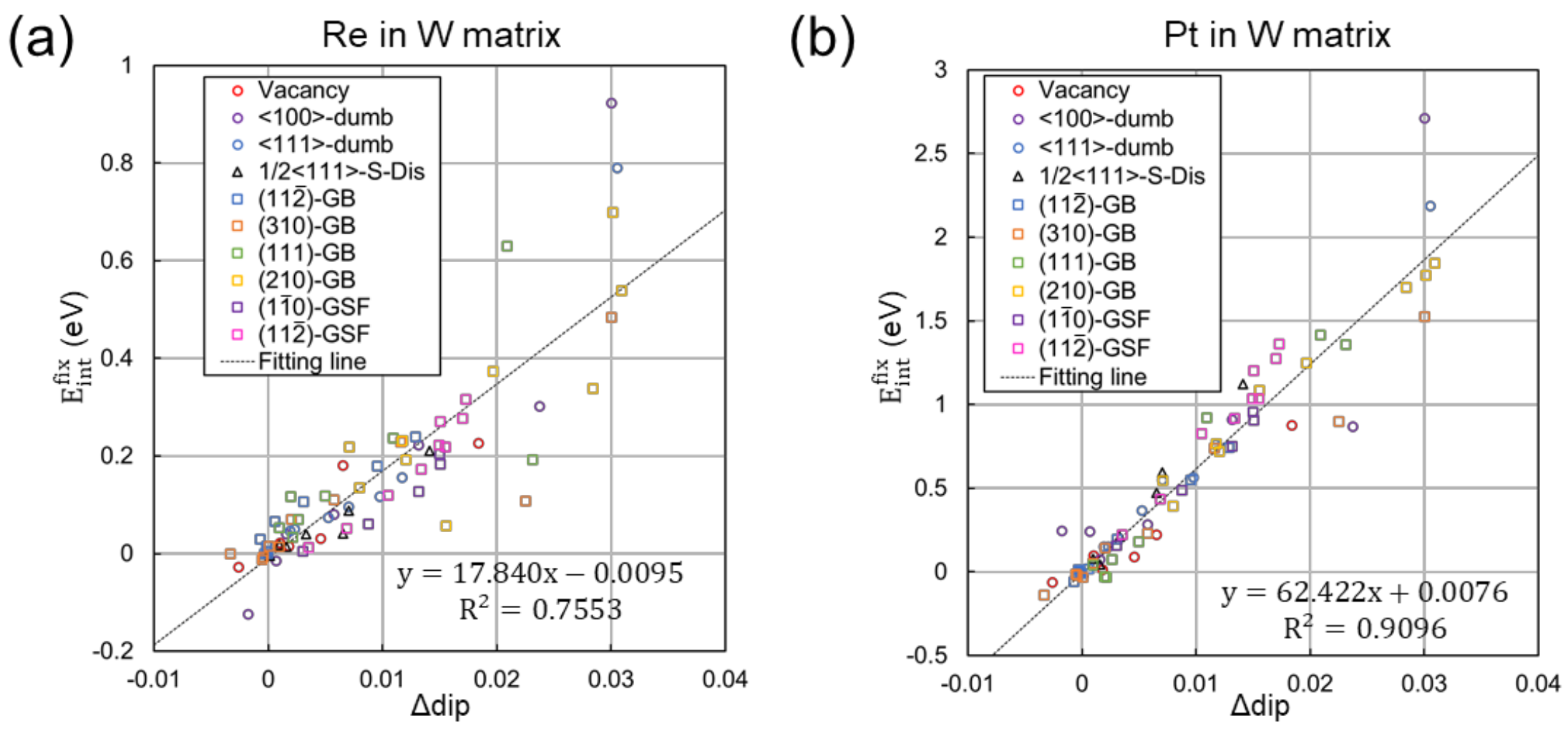

**Figure S7. Linear correlations between $E_{int}^{fix}$ and $\Delta dip$ in the W-Re and W-Pt binary alloys.** The calculated solute-defect interaction energies ($E_{int}^{fix}$) in W are plotted with respect to the corresponding $\Delta dip$ parameters for different atomic sites near various types of defects. (a) Re as the solute; (b) Pt as the solute.

## 7. Hybridization between the valence *sp*- and *d*-bands in bcc W

Besides the importance of the *d*-band, the energy contributions from the valence *sp*-band were found to be crucial in accurately characterizing many fundamental physical properties of transition metal elements, such as cohesive energy[7], equilibrium atomic volume[8,9] and bulk modulus[8,9]. On the one hand, the valence *s*-band strongly overlaps and hybridizes with the valence *p*-band, even though the *p*-band is initially unoccupied in the free-atomic state[10]. On the other hand, both the *s*- and *p*-bands are strongly influenced by the presence of the *d*-band. For example, the valence *sp* electrons may be squeezed into the ion core region due to the large covalent *d*-bonding forces, which results in necessary repulsive forces against the strongly attractive *d*-contributions to maintain the equilibrium atomic distance[8,10]. In addition, it was pointed out that the hybridization between the valence *d*-band and *sp*-band results in large changes in the electronic density of states (DOS), which could provide a significant effect on the cohesive energy in the noble and transition metals[7]. Fig. S8 shows the LDOS of the *s*-, *p*- and *d*-bands of perfect bcc W. As seen from the figure, the peaks of *s*- and *p*-bands are largely overlapped with that of the *d*-bands, which qualitatively indicates a hybridization behavior between the valence *sp*-band and *d*-band.

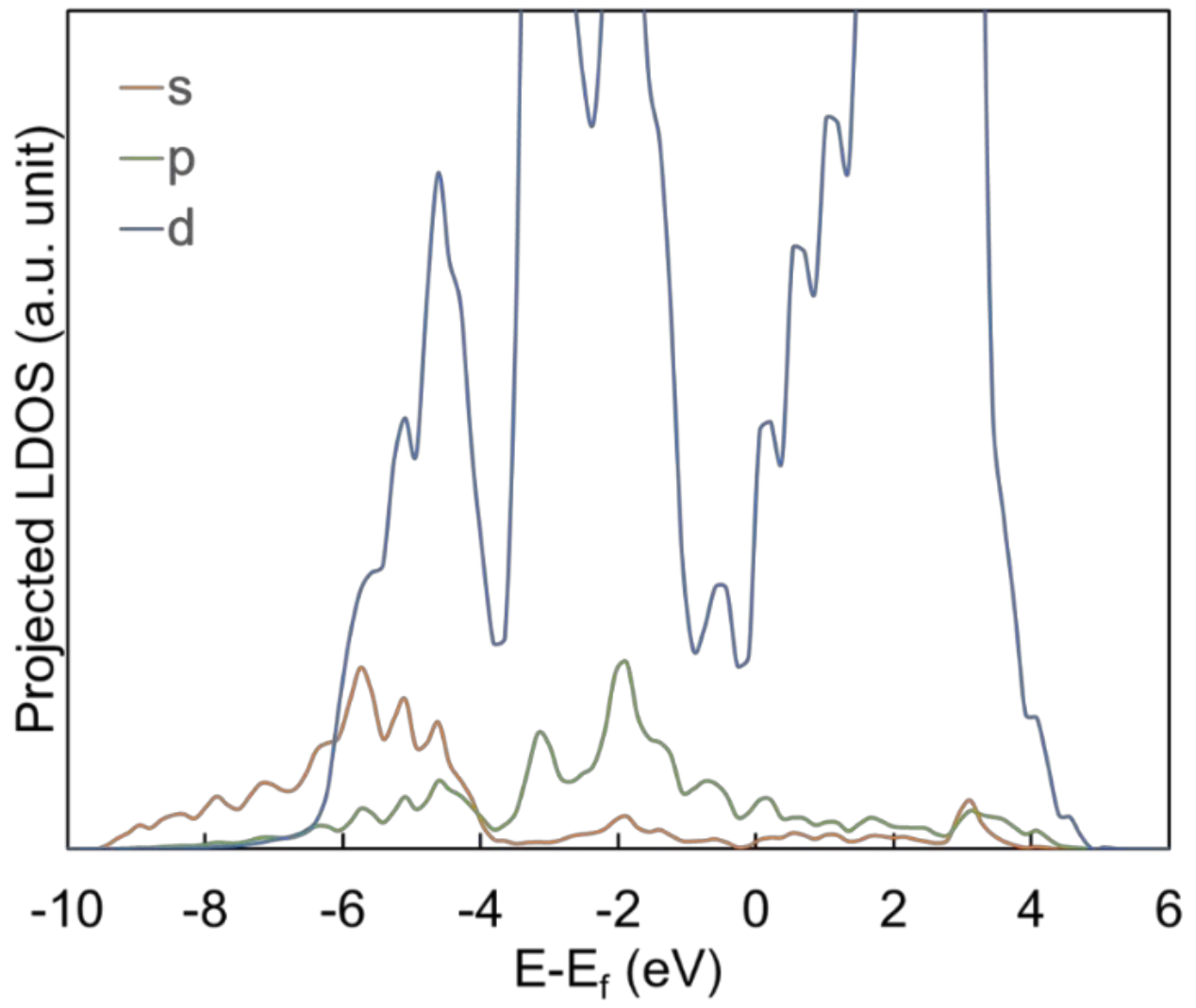

**Figure S8.** Projected LDOS of the *s*-, *p*- and *d*-band of W atoms in perfect bcc lattice with equilibrium lattice parameter. It should be noted that the pseudopotential used for the calculation also treats the semi-core *5p* electrons as valence electrons. However, it is found that the band result from *5p* electrons is localized at very low energy states far away from Fermi level, and thus not plotted here.

Based on the bond-order potential theory, the influence of the *d*-band on *sp*-band can be screened in terms of the local density and environment that can be expressed as a function of interatomic distances[11,12]. To view the variations in the valence *sp*-band with changes of interatomic distances, Fig. S9 shows the LDOS of the valence *6s* and *6p* bands of W atoms in the perfect bcc lattice but with different lattice parameters. In addition, the LDOS of a W atom in unbonded state, which can be considered as the infinite large interatomic distance, is also included for comparison. As seen from Figs. S9a and 9b, before bonding, the LDOSs of both *s*- and *p*-band are single peaks since no hybridization occurs (brown solid-line). Also, the *6p*-band is unoccupied as it is above the Fermi level. As W atoms bonded in a perfect bcc lattice, their *6s*- and *6p*-bands become broad due to the effect of hybridization. In addition, because of the charge transfer caused by hybridization, the initially unoccupied *6p*-band becomes partially occupied. Furthermore, as shown in Fig. S9, the *sp*-band becomes broader when the interatomic distance is shorter (i.e. smaller lattice parameter), implies a stronger hybridization effect. This actually can be anticipated because shorter interatomic distance results in larger *d*-bond integrals, and thus stronger covalent *d*-bonding forces[11,12].

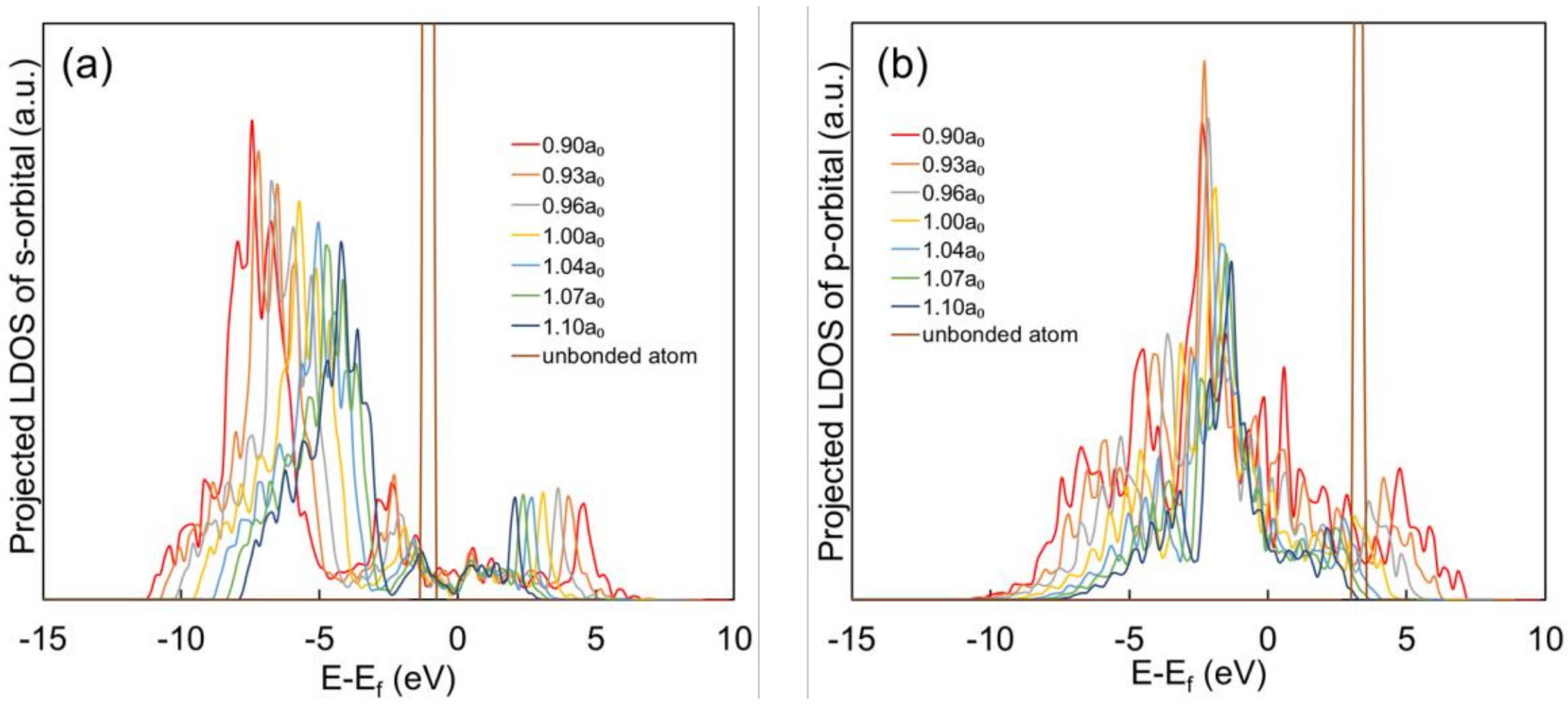

**Figure S9.** The LDOS of of W atoms in the perfect bcc lattice but with different lattice parameters. (a) *6s* band; (b) *6p* band. The length of the equilibrium lattice parameter of bcc W is represented as $a_0$ in the legend.

According to an energy band model developed by Hodges[7], Mueller[13] and Pettifor[8,14], the strength of the hybridization between the valence *sp* and *d* bands in transition metal elements is propositional to the width of the local *d*-band. Furthermore, the local *d*-band width of an atom is closely related to the magnitude of the *d*-*d* interaction matrix elements ($V_{dd}^{ij}$) between the atom and its neighboring atoms[15,16]. The relationship is written as[15,16],

$$W_i \propto \sum_j V_{dd}^{ij} \qquad \text{Eq. S1}$$

where $j$ represents the neighboring atoms of atom $i$, $W_i$ is the local $d$-band width and $V_{dd}^{ij}$ is the *d*-*d* interaction matrix element between $i$ and $j$, which can be scaled as[16,17],

$$V_{dd}^{ij} \propto \frac{r_{d_i}^{\frac{3}{2}} r_{d_j}^{\frac{3}{2}}}{d_{ij}^5} \qquad \text{Eq. S2}$$

where $d_{ij}$ is the interatomic distance between atom $i$ and $j$, and $r_{d_i}$ is the spatial extent of *d*-orbital of atom $i$, which is an intrinsic element-property[16]. Therefore, the strength of the *sp*-*d* hybridization ($E_{sp}$) of an atom $i$ in transition metal alloys can be estimated as,

$$E_{sp} \propto W_i \propto \sum_j V_{dd}^{ij} \propto \sum_j r_{d_i}^{\frac{3}{2}} r_{d_j}^{\frac{3}{2}} / d_{ij}^5. \qquad \text{Eq. S3}$$

The above derivation suggests that the strength of the *sp*-*d* hybridization in a defect structure should be varied with each individual atom since the interatomic distances ($d_{ij}$) of each defect site can be different due to the presence of the defect geometry and the spatial extent of *d*-orbitals of the solute element ($r_{d_i}$) can be different from that of the matrix element. Therefore, the effects of the *sp*-*d* hybridization may not be ignored to understand the solute-defect interactions in the bcc refractory alloys.

## 8. Interactions between the transition metal solutes and $[111](1\bar{1}0)$ GSF in Ta matrix

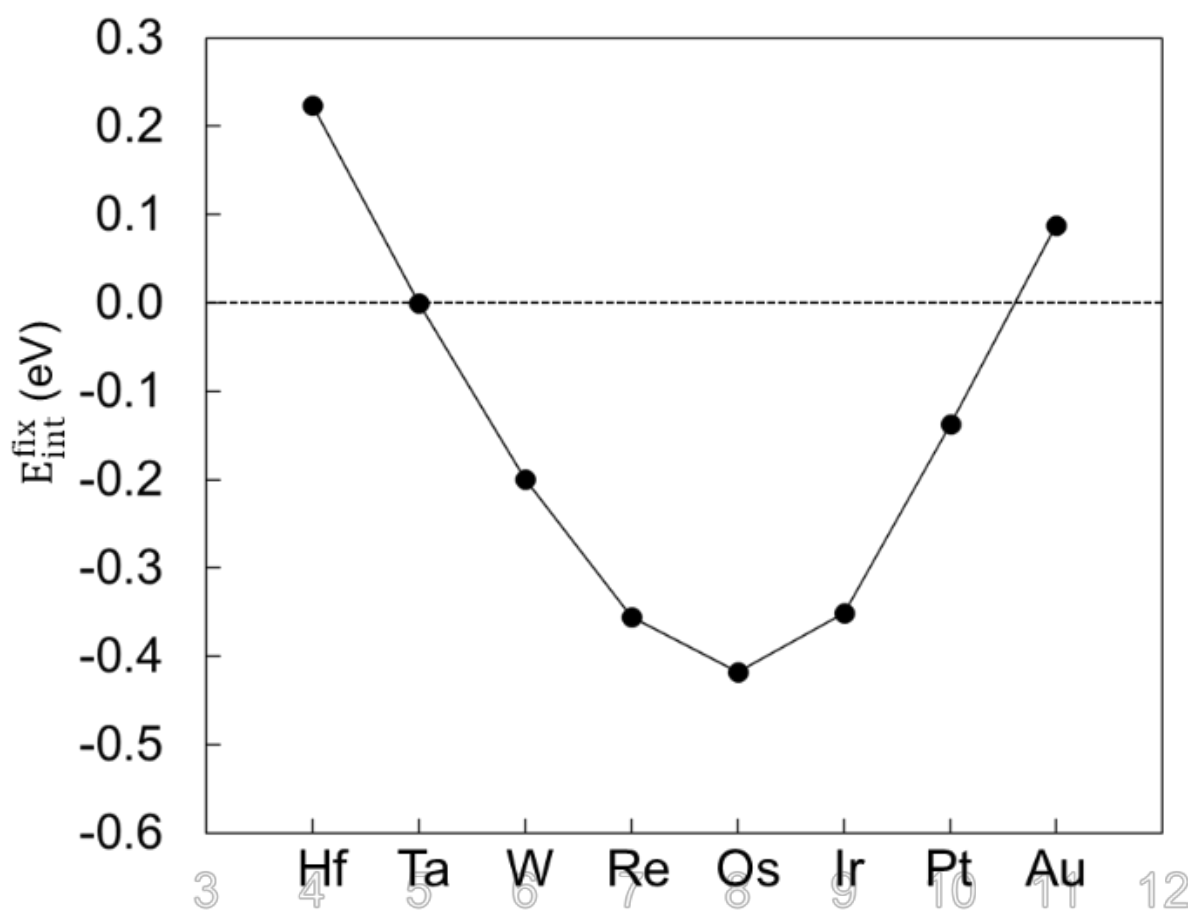

**Figure S10.** Calculated interaction energies of the $(1\bar{1}0)[111]$ GSF with the transition metal solutes in the Ta matrix. The GSF has a relative displacement distance equal to a half of the full Burgers vector. The interaction energies are calculated under the fixed-lattice condition based on optimized GSF structure in pure Ta.

## 9. Electronic dependence of the regression coefficient of the $x_{sp}$ term

As shown in Table 1 in the main text, the regression coefficient of the $x_{sp}$ term, $a_2$, generally has a positive sign if the solute element has fewer *d* electrons than the matrix element (e.g. W-Ta and Ta-Hf), while a negative sign if the difference in the number of d electrons is reversed. This correlation can be understood in terms of the difference in the spatial extent of *d*-orbital between solute and matrix elements. In the present work, the interaction energy is defined as the energy difference between the defect structures with a solute atom occupying the reference site far away and a site near the defect. Therefore, the second part of Eq. 1 ($\Delta E_{sp}$) in the main text can be approximated as an energy difference written as,

$$\Delta E_{sp} \approx \left(E_{sol}^{ref} + E_{mat}^{def}\right) - \left(E_{sol}^{def} + E_{mat}^{ref}\right) \qquad \text{Eq. S4}$$

where $E_{sol}^{def}$ and $E_{sol}^{ref}$ are the energy gain of the solute atom associated with the *sp-d* hybridization while $E_{W}^{def}$ and $E_{sol}^{ref}$ are the energy gain of the matrix element atom. The superscript *def* represents the atom occupying a defect site and *ref* represents that the atom occupying the reference site. Combining Eq. S4 with Eq. S3, we have,

$$\Delta E_{sp} \propto \frac{r_{sol}^{\frac{3}{2}} r_{mat}^{\frac{3}{2}}}{\left(V_{vor}^{ref}\right)^{\frac{5}{3}}} + \frac{r_{mat}^{\frac{3}{2}} r_{mat}^{\frac{3}{2}}}{\left(V_{vor}^{def}\right)^{\frac{5}{3}}} - \frac{r_{sol}^{\frac{3}{2}} r_{mat}^{\frac{3}{2}}}{\left(V_{vor}^{def}\right)^{\frac{5}{3}}} - \frac{r_{mat}^{\frac{3}{2}} r_{mat}^{\frac{3}{2}}}{\left(V_{vor}^{ref}\right)^{\frac{5}{3}}} \qquad \text{Eq. S5}$$

and Eq.S5 can be rewritten as,

$$\Delta E_{sp} \propto r_{mat}^{\frac{3}{2}} \cdot \left(r_{sol}^{\frac{3}{2}} - r_{mat}^{\frac{3}{2}}\right) \cdot \left(V_{vor}^{ref}\right)^{-\frac{5}{3}} \cdot \left(1 - \frac{\left(V_{vor}^{def}\right)^{-\frac{5}{3}}}{\left(V_{vor}^{ref}\right)^{-\frac{5}{3}}}\right) \qquad \text{Eq. S6}$$

Based on Eq. 2 in the main text, if we assume the $\epsilon_{sp}^{def}/\epsilon_{sp}^{ref}$ term usually has a value around 1, the expression on the right side of Eq. S6 can be approximately written as $r_{mat}^{\frac{3}{2}} \cdot \left(r_{sol}^{\frac{3}{2}} - r_{mat}^{\frac{3}{2}}\right) \cdot \left(V_{vor}^{ref}\right)^{-\frac{5}{3}} \cdot x_{sp}$. As a result, the solute element with fewer *d* electrons should have a positive $a_2$ as it also has a more spread spatial extent of *d*-orbital compared to the matrix element[16] (i.e. $r_{sol} > r_{mat}$), and vice versa for the solute element with more *d* electrons than the matrix element.

## 10. Extrapolation ability of the proposed solute-defect interaction model

Here, a linear regression based on Eq. 1 in the main text is performed for a part of interaction energy data in the W-Re system, in which the data of the mono-vacancy and $\frac{1}{2}\langle 111\rangle$ screw dislocation are not included. Then, the obtained regression coefficients are used to predict the interaction energies of the mono-vacancy and $\frac{1}{2}\langle 111\rangle$ screw dislocation and compare with the results from DFT-calculations. As shown by a parity plot in Fig. S11, most of the predictions are in good agreement with the DFT results (RMSE $\approx$ 0.025 eV). This indicates that one can use the correlation relationship established from the data of a few of point and planar defects to estimate the interaction energy of each atomic site for the defects with higher computational complexity, such as the $\frac{1}{2}\langle 111\rangle$ screw dislocation in the present work.

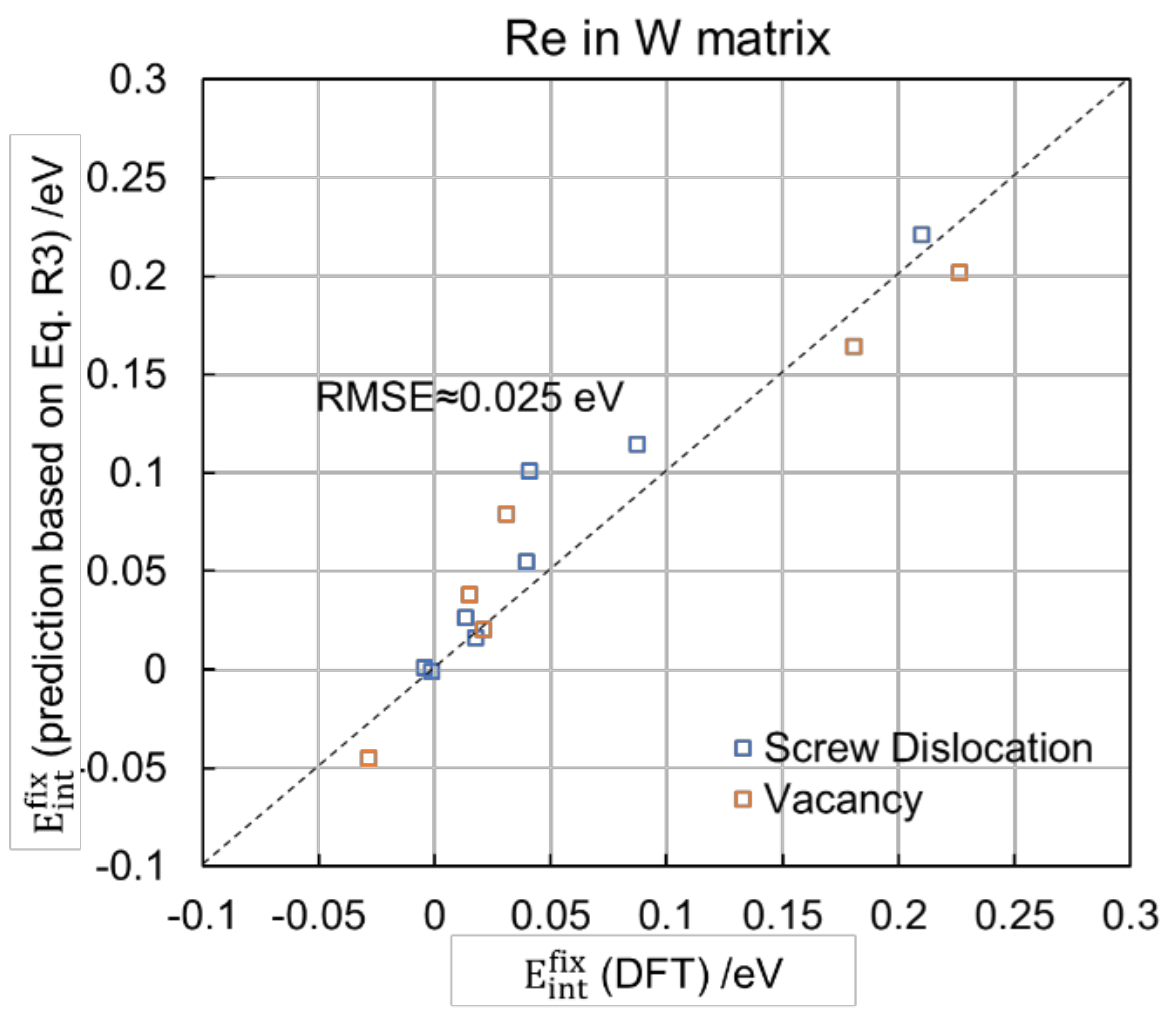

**Figure S11. Comparison between the** model-predicted $E_{int}^{fix}$ of the Re-vacancy and Re-dislocation interactions in W matrix and the results from DFT calculations. The model is established based on Eq. 1 using part of interaction energy data in the W-Re system. In the model regression, the data of the mono-vacancy and $\frac{1}{2}\langle 111\rangle$ screw dislocation were *excluded*.

**Reference:**

1. Kong, X. S. *et al.* First-principles calculations of transition metal-solute interactions with point defects in tungsten. *Acta Materialia* **66,** 172–183 (2014).

2. Bacon, D. J., Barnett, D. M. & Scattergood, R. O. Anisotropic continuum theory of lattice defects. *Progress in Materials Science* **23,** 51–262 (1980).

3. Yasi, J. A. & Trinkle, D. R. Direct calculation of the lattice Green function with arbitrary interactions for general crystals. *Physical Review E* **85,** 66706 (2012).

4. Trinkle, D. R. Lattice Green function for extended defect calculations: Computation and error estimation with long-range forces. *Physical Review B* **78,** 1–11 (2008).

5. Hu, Y.J. *et al.* Solute-induced solid-solution softening and hardening in bcc tungsten. *Acta Materialia* **141,** 304-316 (2017).

6. Wu, X. *et al.* First-principles determination of grain boundary strengthening in tungsten: Dependence on grain boundary structure and metallic radius of solute. *Acta Materialia* **120,** 315–326 (2016).

7. Hodges, L., Ehrenreich, H. & Lang, N. D. Interpolation scheme for band structure of noble and transition metals: ferromagnetism and neutron diffraction in Ni. *Physical Review* **152,** 505 (1966).

8. Pettifor, D. G. Theory of energy bands and related properties of 4d transition metals. III. s and d contributions to the equation of state. *Journal of Physics F: Metal Physics* **8,** 219 (1978).

9. Pettifor, D. G. Theory of energy bands and related properties of 4d transition metals. I. Band parameters and their volume dependence. *Journal of Physics F: Metal Physics* **7,** 613 (1977).

10. Pettifor, D. G. A physicist's view of the energetics of transition metals. *Calphad* **1,** 305–324 (1977).

11. Mrovec, M. *et al.* Bond-order potential for simulations of extended defects in tungsten. *Physical Review B* **75,** 104119 (2007).

12. Znam, S., Nguyen-Manh, D., Pettifor, D. G. & Vitek, V. Atomistic modelling of TiAl I. Bond-order potentials with environmental dependence. *Philosophical Magazine* **83,** 415–438 (2003).

13. Mueller, F. M. Combined interpolation scheme for transition and noble metals. *Physical Review* **153**, 659 (1967).

14. Pettifor, D. G. Accurate resonance-parameter approach to transition-metal band structure. *Physical Review B* **2,** 3031 (1970).

15. Lambert, R. M. & Pacchioni, G. *Chemisorption and Reactivity on Supported Clusters and Thin Films:: Towards an Understanding of Microscopic Processes in Catalysis*. **331,** (Springer Science & Business Media, 2013).

16. Xin, H., Holewinski, A., Schweitzer, N., Nikolla, E. & Linic, S. Electronic structure engineering in heterogeneous catalysis: Identifying novel alloy catalysts based on rapid screening for materials with desired electronic properties. *Topics in Catalysis* **55,** 376–390 (2012).

17. Harrison, W. A. *Electronic structure and the properties of solids: the physics of the chemical bond*. (Courier Corporation, 2012).